\documentclass[twocolumn]{aastex61}

\usepackage{amsmath,amstext}
\usepackage[T1]{fontenc}
\usepackage{apjfonts} 
\DeclareSymbolFont{matha}{OML}{txmi1}{m}{it}

\DeclareMathSymbol{\varv}{\mathord}{matha}{118}

\newcommand\ms{\ifmmode \pazocal{M}_s \else $\pazocal{M}_s$ \fi}

\shorttitle{3PCF as a Probe of the ISM}
\shortauthors{Portillo et al.}

\begin{document}

\title{Developing the 3-Point Correlation Function For the Turbulent Interstellar Medium}

\author{Stephen K.N. Portillo}
\affiliation{Harvard-Smithsonian Center for Astrophysics, 60 Garden Street, Cambridge, MA 02138, USA}

\author{Zachary Slepian}
\affiliation{Einstein Fellow, Lawrence Berkeley National Laboratory, 1 Cyclotron Road, Berkeley, CA 94720, USA}
\affiliation{Berkeley Center for Cosmological Physics, Campbell Hall, University of California at Berkeley, Berkeley, CA 94720, USA}

\author{Blakesley Burkhart}
\affiliation{Harvard-Smithsonian Center for Astrophysics, 60 Garden Street, Cambridge, MA 02138, USA}

\author{Sule Kahraman}
\affiliation{Massachussetts Institute of Technology, Cambridge, MA 02138, USA}

\author{Douglas P. Finkbeiner}
\affiliation{Harvard-Smithsonian Center for Astrophysics, 60 Garden Street, Cambridge, MA 02138, USA}

\begin{abstract}
We present the first application of the angle-dependent 3-Point Correlation Function (3PCF) to the density fields magnetohydrodynamic (MHD) turbulence simulations intended to model interstellar (ISM) turbulence. Previous work has demonstrated that the angle-averaged bispectrum, the 3PCF's Fourier-space analog, is sensitive to the sonic and Alfv\'enic Mach numbers of turbulence. Here we show that introducing angular information via multipole moments with respect to the triangle opening angle offers considerable additional discriminatory power on these parameters. We exploit a fast, order $N_{\rm g} \log N_{\rm g}$ ($N_{\rm g}$ the number of grid cells used for a Fourier Transform) 3PCF algorithm to study a suite of MHD turbulence simulations with 10 different combinations of sonic and Alfv\'enic Mach numbers over a range from sub to super-sonic and sub to super-Alfv\'{e}nic.  The 3PCF algorithm's speed for the first time enables full quantification of the time-variation of our signal: we study 9 timeslices for each condition, demonstrating that the 3PCF is sufficiently time-stable to be used as an ISM diagnostic. In future, applying this framework to 3-D dust maps will enable better treatment of dust as a cosmological foreground as well as reveal conditions in the ISM that shape star formation.
\end{abstract}


\section{Introduction}
\label{sec:intro}
Dust \added{and gas} in the interstellar medium (ISM) \replaced{plays}{play} a critical role in Galactic evolution.  \added{In particular,} \replaced{the}{The} dust provides cooling pathways for dissipating energy, which aids clouds in becoming gravitationally self-bound \citep{Omukai00a,Schneider2003}. Furthermore, ISM dust in emission or absorption is a good tracer of the overall density in the ISM \citep{goodman09}. Dust also acts as a foreground for most extragalactic measurements, from those of the Cosmic Microwave Background (CMB) temperature and polarization fluctuations to the reddening of stars, galaxies, and transients such as supernovae or gamma ray bursts \citep{PlanckXLIV,PlanckXXIX,PlanckL,Cardelli89,Corasaniti06}. It can be a particularly important foreground in searches for inflationary signatures, both via tensor modes produced by primordial gravity waves (``B-modes'') and for scale-dependent bias \citep{Dalal08} on large scales in the power spectrum of galaxy redshift surveys. 

Quantifying the spatial structure of interstellar gas and dust in our own Galaxy is thus important both in enabling more predictive models of planet and star formation and for removal of extragalactic foregrounds. Complicating this picture, however, is that the interstellar medium (ISM) is highly turbulent and magnetized \citep{ElmegreenScalo,Mckee_Ostriker2007}. 
Turbulence produces gas and dust density fluctuations via an energy cascade that may extend over scales from kiloparsecs down to the proton gyroradius, as is evident from the line-width-size relation \citep{Larson81a}, the ``big powerlaw'' of electron scintillation and scattering measurements \citep{Armstrong95,CheL10} and the power spectrum of 21-cm gas \citep{burkhart10, chepurnov15}.
Despite recent theoretical advances in understanding compressible and incompressible MHD turbulence \citep{Goldreich95a,cho03,Boldyrev,lazarianberesnyak06, Kowal07a,Kowal10}, ISM turbulence remains analytically intractable due to complex injection sources, different phases and various
instabilities acting at multiple scales. Furthermore, the basic plasma turbulence properties of the medium, such as the sonic and Alfv\'enic Mach numbers, the injection scale or scales, and the dissipation mechanisms are largely unknown due to observational issues such as line-of-sight integration, radiative transfer effects and telescope beam smearing.

The difficulties mentioned above mean that numerical
simulations coupled with statistical diagnostics are necessary to investigate turbulent density structures
in the ISM quantitatively. Present MHD turbulence simulations can reproduce many observed dust structures, such as filaments \added{\citep{2016MNRAS.457..375F}} and fractals \added{\citep{Federrath09a,burkhart13a}}. However because of their limited numerical resolution, simulations cannot reach the observed Reynolds numbers (i.e. the ratio of inertial to viscous forces) of the ISM. Nonetheless, simulations represent fairly well the statistical properties of the ISM, including the probability distribution function of density \citep{vazquezsemadeni97,Federrath09a, Burkhart09,Kainulainen09a,Burkhart12, burkhart15,chen2017}, velocity and density power spectra \citep{Stanimirovic99a,Stanimirovic04a,LP08,burkhart10,Bur13,chepurnov15,doi:10.1093/mnras/stt1644}\explain{Added reference to Federath 2013}, and principal components \citep{Heyer04a,Hey12}. 
Statistical studies are therefore one promising avenue to connect theory, simulation, and observation of turbulence in the ISM.
Other statistical techniques include higher-order-spectra, such as the bispectrum \citep{Burkhart09,burkhart10,Cho2009,Burkhart2016ApJ...827...26B}, higher-order moments \citep{Kowal07,burkhart10,Gaensler2011}, topological
techniques \citep{Kowal07,Chepurnov2008}, clump and hierarchical structure finders \citep{Rosolowsky08a, goodman09,burkhart13a}, Tsallis
functions \citep{Esquivel11,tofflemire11}, and structure or correlation
functions as tests of intermittency and
anisotropy \citep{CL03,esquivel05,KowL10,PhysRevLett.101.194505,2012JFM...692..183K}\explain{references to Schmidt+ and Konstandin+ added}.

The most commonly-used statistical tool of turbulence is the spatial or temporal Fourier power spectrum. The power spectrum is useful for studying the turbulent power cascade as a function of scale or frequency and searching for the sources and sinks of energy, also known as the injection and dissipation scales \citep{LP04,LP06,Kowal07,0004-637X-665-1-416,goodman09,Heyer09a,10.1051/0004-6361,burkhart10,2010A&A...512A..81F,Collins12a,burkhart14,chepurnov15}\explain{references to Kritsuk+, Schmidt+, Federrath+ added}.
However a critical limitation of the power spectrum is that it ignores phase information that is important for capturing the overall structure.  In order to study both phase and amplitude information a number of statistics beyond the power spectrum have been proposed for ISM studies, including the bispectrum \citep{Burkhart09,burkhart15,Burkhart2016ApJ...827...26B}, which studies the correlations of triplets of the Fourier-transformed density evaluated at three wavenumbers that form a closed triangle. The configuration-space analog of the bispectrum is the 3-Point Correlation Function (3PCF), which probes excess triangles (relative to a random distribution) formed by density points.  The 3PCF is a natural tool for turbulence studies as it encodes information about non-linear interactions that is absent in the 2-Point Correlation Function (2PCF), the configuration-space analog of the power spectrum.

However, a difficulty faced by bispectrum and 3PCF studies is that direct calculations are numerically expensive as the choice of wave-vector triplets scales as $N^3$, with $N$ the number of wavevector magnitudes: one requires three to specify a Fourier-space triangle. Previous work somewhat evaded this difficulty by averaging over the triangle opening angle in Fourier space, reducing the scaling to $N^2$ as there remain only two sides as free parameters \citep{Burkhart09}.

Recently \citet{Slepian15_alg} introduced a fast 3PCF algorithm that reduced the computational expense of the calculation from  $N^3$, with $N$ the number of points, to $N^2$ using spherical harmonics, and further accelerated it to $N_{\rm g}\log N_{\rm g}$ using Fourier Transforms \citep{Slepian16_alg_WFTs}, with $N_{\rm g}$ the number of grid cells used for the Fourier Transform. Furthermore, the algorithm allows for calculation of higher multipole moments and does not use angle averaging as was done in \citet{Burkhart09}. The $N^2$ version of the algorithm (spherical-harmonic-only) has been successfully applied to large-scale structure (\citealt{Slepian16_comp_3PCF, Slepian16_RV_const, Slepian17_BAO_3PCF}), but the current work is the first application of the Fourier-Transform-based version.

Here, we use the Fourier-based algorithm to perform the first study of the angle-dependent 3PCF for simulations of the turbulent, magnetized ISM. We focus on how sensitive the 3PCF multipole moments are to the sonic and Alfv\'enic Mach numbers. These are important turbulence parameters defined as respectively the sonic and Alfv\'enic Mach numbers $M_s \equiv  |\pmb{\varv}|/c_s$ and $M_A \equiv |\pmb{\varv}|/ \langle \varv_A \rangle$. $\pmb{\varv}$ is the velocity, and $c_s$  and $\varv_A$ are respectively the isothermal sound speed and the Alfv\'en speed.

The paper is structured as follows. In \S\ref{sec:basis_alg} we review the basis and algorithm and outline the numerical implementation enabling the high speed necessary for this study. \S\ref{sec:analysis_tech} presents two compressions of the 3PCF we perform as well as a quantitative fitting ansatz for them. \S\ref{sec:sims} describes the MHD turbulence simulations we use. \S\ref{sec:results} discusses our results, and \S\ref{sec:concs} concludes.

\begin{figure}
\includegraphics[width=1\columnwidth]
{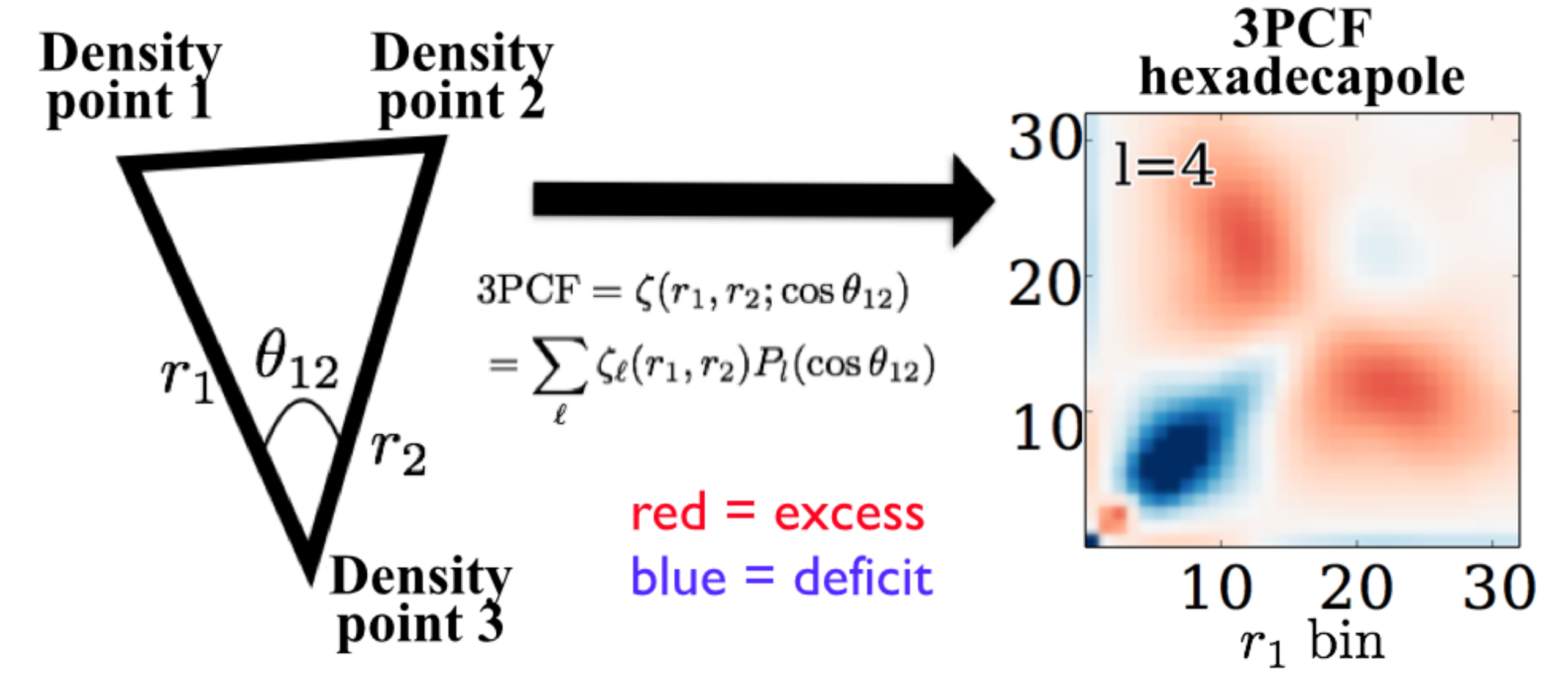}
\caption{Left: the geometry for the 3PCF is shown. We parametrize it by two triangle sides $r_1$ and $r_2$ and their enclosed angle $\theta_{12}$, expanding the full 3PCF in a Legendre series with radial coefficients $\zeta_{\ell}(r_1, r_2)$ for the dependence on side length times Legendre polynomials $P_{\ell}$ for the angular dependence. These coefficients can then be plotted versus binned $r_1$ and $r_2$, where red represents an excess of triangles over random and blue a deficit; $\ell = 4$ is shown as an example at right. Here each bin corresponds to 4 simulation pixels.}
\label{fig:alg_fig}
\end{figure}

\section{Basis \& Algorithm}
\label{sec:basis_alg}
\subsection{Multipole expansion}
One can imagine the 3PCF in terms of points and the excess frequency of triangles of a certain shape whose vertices are given by those points, or alternatively, the excess product of density in triples of grid cells at the corners of triangles of a certain shape.  We begin our discussion by considering the former, and then derive the latter.  Throughout this work we measure the full 3PCF of the simulations using the Legendre basis. We expand the 3PCF as a sum over Legendre polynomials (multipoles) containing the dependence on the opening angle of the triangle, weighted by radial coefficients describing the dependence on triangle side length. Mathematically we have
\begin{align}
\zeta(r_1, r_2;\hat{r}_1\cdot\hat{r}_2)= \sum_{\ell} \zeta_{\ell}(r_1, r_2)P_{\ell}(\hat{r}_1\cdot\hat{r}_2),
\label{eqn:3pcf_expansion}
\end{align}
with $\zeta$ the full 3PCF, $\zeta_{\ell}$ the multipole moments, and $P_{\ell}$ the Legendre polynomials. $\hat{r}_1\cdot\hat{r}_2 =\cos \theta_{12}$ and $\theta_{12}$ is the opening angle of the triangle. Figure \ref{fig:alg_fig} summarizes the geometry and basis and shows an example of the ultimate result, the coefficient $\zeta_{\ell}(r_1, r_2)$. If the sum over $\ell$ ranges from zero to infinity, the Legendre polynomials form a complete (and orthogonal, but not orthonormal) basis for an arbitrary 3PCF of $r_1, r_2$, and $\hat{r}_1\cdot\hat{r}_2$. However, in practice we truncate the sum at some finite $\ell_{\rm max}$; here we choose $\ell_{\rm max}=5$. This likely leaves some information about the angular dependence unexplored, though we can check the convergence of this truncated expansion by forming cumulative sums over all multipoles up to and including a given $l$. In practice we find that for the raw 3PCF this cumulative sum is dominated by the first few $\ell$, meaning that the full 3PCF is a relatively smoothly-varying function of triangle opening angle. 

The statistic we choose to report in this work is actually the 3PCF multipoles divided by their standard deviation in time, and these cannot be directly added to obtain the full raw 3PCF, nor to obtain the full 3PCF divided by the full noise. However, they are the most useful statistic to examine because each multipole has similar signal to noise: for instance, while the raw monopole is the largest amplitude, it also varies the most in time. The convergence of the information in these signal-to-noise weighted moments is evident because the higher moments begin to look similar to each other (see Figures \ref{fig:MA7_full} and \ref{fig:MA_pt7_full}).

The algorithm proceeds by obtaining local estimates of the multipole coefficients about a given center $\vec{x}$ and then averaging over all $\vec{x}$ in the end to yield the translation-averaged 3PCF. The local estimate of the multipole coefficients can be evaluated using orthogonality, such that about a given point $\vec{x}$ we integrate over all opening angles, parametrized by $\mu_{12}\equiv\cos \theta _{12}=\hat{r}_1\cdot\hat{r}_2$. We have
\begin{align}
\zeta_l(r_1, r_2;\vec{x}) =\frac{2l+1}{2} \int d\mu_{12}\;P_{\ell}(\mu_{12})\zeta(r_1,r_2;\mu_{12};\vec{x}),
\label{eqn:zeta_ell}
\end{align}
where $\zeta(r_1, r_2;\mu_{12};\vec{x}) = \left<\delta(\vec{x})\delta(\vec{x}+\vec{r}_1)\delta(\vec{x}+\vec{r}_2)\right>$ and the angle brackets denote averaging over rotations. $\delta$ is the density field, and $\delta(\vec{x})$ comes outside the integral and can be applied as a weight once the integral has been performed. The pre-factor is to deal with the fact that the Legendre polynomials are orthogonal but not orthonormal. 

As shown in \cite{Slepian15_alg}, the integral can be evaluated by rewriting the Legendre polynomial as a product of spherical harmonics, one in $\hat{r}_1$ and one in $\hat{r}_2$, via the spherical harmonic addition theorem. We will then integrate over $d\Omega_1d\Omega_2$, which takes care of averaging over rotations as well; thus we can remove the angle brackets when the integration is written in this form. The integral over $\mu_{12}$ is now separable in $d\Omega_1d\Omega_2$. Using spherical harmonics means that we can compute the local estimate of the multipole moments of the 3PCF by assembling local spherical harmonic coefficients on radial bins about each point $\vec{x}$\replaced{ .}{. }Around a given point $\vec{x}$, one thus has an order $N$ calculation, with $N$ the number of points, and one must do this calculation about each $\vec{x}$, leading to an overall $N^2$ scaling. 

Obtaining the harmonic coefficients about all $\vec{x}$ can be recast as a convolution and therefore evaluated with Fast Fourier Transforms (FFTs) if the density field is gridded, as shown in \cite{Slepian16_alg_WFTs}. This reduces the computational cost of the algorithm from $N^2$ to $N_{\rm g}\log N_{\rm g}$, with $N_{\rm g}$ the number of grid cells used for the FFT.

While using the FFT version of the algorithm demands a trade-off between speed and precision for data that is not initially gridded, for initially gridded data the FFT version of the algorithm is exact. This is the case for the simulations we use in this work, which are all performed on a regular, periodic cubic grid.

\begin{figure*}
\includegraphics[scale=0.9]{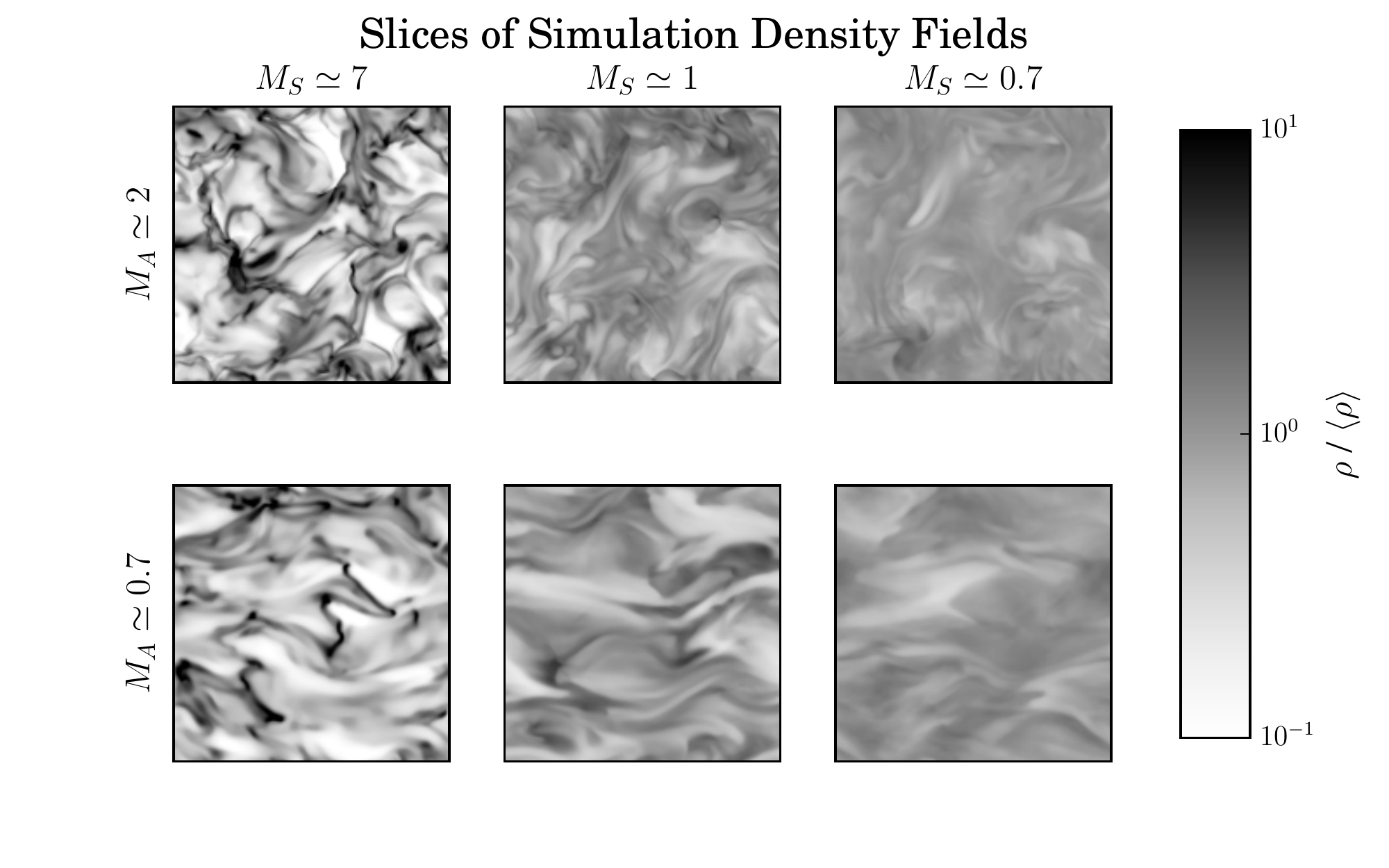}
\caption{Density field slices of a representative range of the simulations used in this work, shown using a logarithmic color map. The high density contrast from shocks in the supersonic simulations can clearly be seen, as well as finer structures arising in the super-Alfv\'enic simulations. These slices are the raw density field $\rho$ divided by the average density $<\rho>$, not the fluctuation field $\delta$ of which we compute the 3PCF.}
\label{fig:sim_boxes}
\end{figure*}

\subsection{Implementation of the FFT-based 3PCF algorithm}
\label{sec:imp}
Here we briefly outline our implementation of the FFT-based 3PCF algorithm. First, the algorithm convolves the gridded spherical harmonic kernel for a given $\ell$ and $m$ on a given radial bin with the density field to generate the harmonic expansion coefficient $a_{\ell m}$ on that bin about every possible origin in the box simultaneously. 

At fixed data cube size and choice of radial bins, the spherical harmonic kernels are independent of the density data. The FFTs of these kernels are pre-computed and stored on disk so that they can be re-used in calculating the 3PCFs of different simulations.

The density cube is read from disk and its FFT computed only once. To ensure that the problem fits in memory, we consider the spherical harmonic kernels one at a time. We load from disk the FFT of the kernel for a given $\ell$, $m$, and radial bin and then multiply it by the FFT of the data. The inverse FFT of this product then yields the desired harmonic expansion coefficient grid, which is saved to disk. This procedure requires three arrays of the box size in memory at once: the kernel FFT, the data FFT, and their product. For a $256^3$ simulation, each double-precision array of complex numbers is about $256^3\times 2\times 8{\;\rm bytes}=268$ MB, leading to $3\times 268{\;\rm MB}=0.81$ GB total; for $512^3$ the total memory requirement is then 6.44 GB, and for $1024^3$ it is 51.54 GB. Working in double precision is important when taking the FFT.

Once all the harmonic coefficients are obtained, we must form all possible combinations of two different radial bins at the same $\ell$ and $m$, multiply them, sum over spins $m$ and then weight by the real-space density field\deleted{.} as discussed after equation (\ref{eqn:zeta_ell}). The harmonic coefficients are read from disk for each pair, so this step has three box-sized grids in memory at the same time: the two harmonic coefficients entering the bin combination and the accumulating sum over $m$. Once the sum over $m$ has been computed, one can then load the real-space density field, apply it as a final weight, and sum over $\vec{x}$, which is equivalent to translation-averaging.

The approach above, in which only three arrays are loaded, is appropriate for a machine with limited memory. However another approach can be taken when more memory is available. In this case, it is more efficient to load the harmonic coefficient grids on all bins at once as for a given $\ell$ and $m$, there is significant reuse of each radial bin. The coefficient grid for a given radial bin will be paired with that for all other radial bins, meaning it is not ideal to read in the grids needed for one combination of bins, only to shunt one or both out of memory for the next combination. This strategy enables a performance tradeoff where computational speed can be gained at the cost of increased memory requirements. For $256^3$ boxes, it was feasible to hold harmonic coefficient grids for 32 different bins on the nodes we used (which have 4 AMD Opteron Abu Dhabi CPUs with 64 GB of RAM each), but larger simulations would exceed the memory we had available.

\added{When calculating the 3PCF of very high-resolution data sets, like the $4096^3$ simulations in \cite{doi:10.1093/mnras/stt1644}, the FFTs to compute the harmonic expansion coefficient grids will become the rate-limiting step, scaling as $N_g \log_2 N_g$. To compute the harmonic expansion coefficient grids for $256^3$ boxes, up to $\ell=5$ for 32 radial bins, requires 672 FFTs at 13 seconds each, or 2.5 core-hours total. In this work, we have not parallelized our code, opting instead to have each CPU calculate the 3PCF of a different data set. Thus, for a $4096^3$ dataset which has $(4096/256)^3 = 4096$ times as many voxels \added{as in our $256^3$ runs}, we expect each FFT to take $4096 \log_2 4096 = 49152$ times longer, or 7 days each for a total of 14 core-years. Clearly, this calculation would need to be parallelized to be feasible. The FFTs are independent, since each is obtaining the harmonic coefficient for a particular $\ell$ and $m$ and on a single, particular radial bin. They could thus easily be distributed among cores. If parallelized in this way, a $4096^3$ grid would take about 7 days with 672 cores. The FFTs themselves could also be parallelized by domain decomposition. For example, the PFFT implementation \citep{10.1137/120885887} allows for a 2D domain decomposition and has been demonstrated on $8192^3$ grids using up to 262,144 cores
on a BlueGene/P architecture. The memory requirements of calculating a $4096^3$ 3PCF would be substantial, requiring 3.30 TB. Parallelizing the computation may increase the total memory requirement, but would allow the memory to be distributed over many cores.}

\section{Analysis}
\label{sec:analysis_tech}
The density fields of turbulent media are often log-normal, meaning that the logarithm of the density fluctuations follows a Gaussian distribution \citep{Vazquez-Semadeni1994,Federrath2008,Burkhart2012}. All the information for Gaussian random fields is contained in the 2PCF or power spectrum, and if the turbulent density field is only weakly non-Gaussian in log space, this suggests that the 2PCF and 3PCF of its logarithm will capture all the information. The case is analogous to cosmology, where on large scales the density field is only weakly non-Gaussian due to late-time non-linear structure formation altering the initially Gaussian random density. Thus on large scales in cosmology the 2PCF and 3PCF are taken to nearly fully characterize galaxies' spatial clustering on scales of tens to hundreds of Megaparsecs.

We therefore compute \deleted{an }a density fluctuation field for the scaled logarithm of our simulated density cubes, as
\begin{align}
\delta = \frac{\log \rho - \left<\log\rho\right>}{\sigma(\log\rho)},
\end{align}
where the base of the logarithm does not matter as it cancels out in this ratio. The density fluctuation field for each simulation cube thus has zero mean and unit variance. We then take the 3PCF of this density fluctuation field. This choice normalizes for the scale of density fluctuations in each simulation, allowing us to focus on the differences in non-Gaussian structure between simulations. The power spectrum or 2PCF will already capture differences in the overall variance, as the variance is the zero lag limit of the 2PCF or the integral of the power spectrum over all $k$. 

Once we have the multipole moments of the 3PCF as a set of functions of $r_1$ and $r_2$, we need a means of relating them to the sonic and Alfv\'enic Mach numbers used for the simulation. We perform two empirically-motivated compressions of the data which we describe below. A related compression scheme is described in \citet{Burkhart2016ApJ...827...26B}. We then fit to them a simple form that models the 3PCF dependence on these parameters with a power law in each, with an additional constant pre-factor that depends only on $\ell$.

The first compression we form is to consider the 3PCF along two lines: the diagonal $r_2 = r_1$ and the half-diagonal $r_2 = r_1/2$. This compression leads to a 1-D function of $r_1$ at each multipole $l$. We will have such a function for each pair of $M_S$ and $M_A$ simulated. Thus the dimension of this dataset is $N_{M_S} N_{M_A} (\ell_{\rm max}+1)N_r$, where $N_{M_S}$ and $N_{M_A}$ are respectively the number of $M_s$ and $M_A$ values simulated and $N_r$ is the number of bins in side length we use.

The second compression is to then average these first compressions over all scales, so that we now have a function only of $l$. Again, we will have such a function for each pair of $M_S$ and $M_A$ simulated. The dimension of this dataset is then $N_{M_S} N_{M_A} (\ell_{\rm max}+1)$. We display the results of each of these two compressions in \S\ref{sec:results}.

We now make the ansatz that the fully compressed 3PCF has the form
\begin{align}
\zeta = f(l)\left[c_P \log M_S + c_B \log M_A \right]
\end{align}
where $c_P$ and $c_B$ are constants, with subscripts ``P" for pressure, which enters the sonic Mach number, and ``B" for magnetic field, which enters the Alfv\'enic Mach number.

$c_P$ and $c_B$ can be interpreted as logarithmic derivatives of the 3PCF with respect to $M_S$ and $M_A$, evaluated at some fiducial $M_{S,0}$ and $M_{A,0}$ or equivalently pressure and magnetic field. In other words, we are essentially considering coefficients in a Taylor expansion of the 3PCF as
\begin{align}
\zeta(M_S, M_A) &= \zeta(M_{S,0}, M_{A,0}) + \frac{\partial \zeta}{\partial \log M_S}\bigg|_{M_{S,0}, M_{A,0}}\left[\log M_S - \log M_{S,0} \right]\nonumber\\
&+ \frac{\partial \zeta}{\partial \log M_A}\bigg|_{M_{S,0}, M_{A,0}}\left[\log M_A - \log M_{A,0} \right]\nonumber\\
&+ \mathcal{O}\left(\left(\log M_S - \log M_{S,0}\right)^2\right)\nonumber\\
&+ \mathcal{O}\left(\left(\log M_A - \log M_{A,0}\right)^2\right),
\label{eqn:taylor}
\end{align}
with $c_P \equiv \partial \zeta/\partial \log M_S$ and $c_B =\partial \zeta/\partial \log M_A$. These coefficients are roughly analogous to bias parameters in cosmology, where the galaxy density field is expanded in terms of powers of the matter density field with the biases as fitted coefficients encapsulating the complex physics of galaxy formation (e.g. \citealt{Desjacques16}).

In this paper, one focus will be obtaining the coefficients $c_P$ and $c_B$ from simulations. However, we briefly discuss how once this has been done the process could be reversed to determine the $M_S$ and $M_A$ from data in a given region of the ISM. In this latter case, one can measure the 3PCF, and simulate fiducial values $M_S$ and $M_A$ believed to be close to the true values. One could then solve the system of equations implied by the Taylor series expansion (\ref{eqn:taylor}) to find $M_s$ and $M_A$ for that region of the ISM. Generically the inversion of equation (\ref{eqn:taylor}) is the solution of the $\chi^2$ minimization problem $\chi^2 = \left( \vec{D}-\vec{M}\right) {\bf C}^{-1} \left(\vec{D}-\vec{M}\right)^T$, where $\vec{D}$ is the data vector, $\vec{M}$ is the model vector, ${\bf C}$ is the covariance matrix, which encodes the (presumed) independence structure of the data, and $T$ means ``transpose.''
This minimization can be performed efficiently using matrix algebra.

We now outline possible approaches to the covariance matrix. The simplest approach, which we adopt in this work, is to take it that all pixels in the $(r_1, r_2)$ plane and at each $\ell$ are independent, and to use the time-variation as a measure of the ``cosmic-variance'' standard deviation by invoking the ergodic hypothesis. This  hypothesis implies that averaging over many different time-slices is the same as averaging over many different regions of space. This is exactly the variance we would expect if we measured the 3PCF averaged over many different patches of the ISM, as would be the likely use-case.

However this approach will fail to capture the fact that the pixels in $\ell$, $(r_1, r_2)$ are likely not fully independent of each other. To make a less simplistic treatment of the covariance, we can notice that since the density field is roughly log-normal, computing the 3PCF of its logarithm means that the six-point function (which gives the covariance of the 3PCF) will have a Gaussian random field (GRF), or disconnected, contribution.  Since the field is nearly log-normal, the distribution of logarithms is nearly Gaussian. We thus expect that the GRF piece of the six-point function dominates the other pieces and so is a good estimate of the full 3PCF covariance. The GRF covariance is given by the inverse FT of the product $(P(k_1) + 1/n)(P(k_2) + 1/n)(P(k_3) + 1/n)$, where the $k_i$ form a closed triangle, $P(k)$ is the power spectrum, and $n$ is the effective number density of objects ($1/n$ is the ``shot noise''). One can see that there will be a pure signal contribution scaling as $P^3$, as well as signal-cross-noise contributions $P^2/n$ and $P/n^2$, and a pure noise contribution $1/n^3$. The approach we have adopted in this work captures only the pure noise contribution, which will give only diagonal contributions to a covariance matrix in configuration space (the inverse FT of a constant is a Dirac Delta function). However generically there will also be other contributions to the diagonal not captured, as well as off-diagonal pieces, from the pure signal and signal-cross-noise terms.

The GRF six-point function in the multipole basis was computed in \cite{Slepian15_alg}, where the effective volume $V$ and effective survey number density $n$ (giving the shot noise, $1/n$) were left as free parameters to be fit using a noisy empirical covariance matrix derived from a large number of mock catalogs, a procedure followed through in \cite{Slepian16_comp_3PCF,Slepian17_BAO_3PCF}. Here we do not have a large number of mocks because we do not know the correct underlying $M_S$ and $M_A$ values with which to produce them, so we need to estimate $V$ and $n$ directly. A good first estimate will be the physical volume and number density of the simulation box (or observed volume) used. In the case of galaxies, the effective $V$ and $n$ take on values different from these because they are absorbing some non-linearity that makes the six-point function deviate from its disconnected piece. However, they are not greatly different from the true values. Given that the log-density field should be roughly Gaussian in the present case as well, we expect this will still hold true.

\section{Numerical Simulations of MHD Turbulence}
\label{sec:sims}

We now describe the MHD simulations used to obtain \deleted{realistic} density profiles for sub- and supersonic turbulent gas with different magnetic field strengths.
Our simulations are isothermal and non-self-gravitating.
We use a third-order-accurate hybrid essentially non-oscillatory scheme \citep{Cho2002,Cho2003} to solve the ideal MHD equations,
\begin{align}
 \frac{\partial \rho}{\partial t} + \nabla \cdot (\rho \pmb{\varv}) &= 0, \\
 \frac{\partial (\rho \pmb{\varv})}{\partial t} + \nabla \cdot \left[ \rho \pmb{\varv} \pmb{\varv} + \left( p + \frac{B^2}{8 \pi} \right) {\bf I} - \frac{1}{4 \pi}{\bf B}{\bf B} \right] &= {\bf f},  \\
 \frac{\partial {\bf B}}{\partial t} - \nabla \times (\pmb{\varv} \times{\bf B}) &= 0 \ ,
\end{align}
where $\rho$ is the density,
${\bf B}$ is the magnetic field, $p$ is the gas pressure, and ${\bf I}$ is the identity matrix. We assume the zero-divergence condition $\nabla \cdot {\bf B} = 0$,  periodic boundary conditions, and an isothermal equation of state $p = c_s^2 \rho$, with $c_s$ the isothermal sound speed. The ${\bf B}{\bf B}$ term is the magnetic stress matrix and can also be written as ${\bf B} \cdot \nabla{\bf B}/(4\pi)$. For the source term $\bf{f}$, we assume a random large-scale solenoidal driving at a wave number $k\approx 2.5$ (i.e.~1/2.5 the box size). 
The simulations have $256^3$ resolution elements, and the code has been both employed and described in many previous works
\citep{Cho2003,Kowal2007,Burkhart09, Kowal2009, burkhart10, Kowal2011}. The simulation runs used in this work are publicly available as part of the Catalog for Astrophysical Turbulence Simulations (CATS, Burkhart et al. 2017, in prep.; Burkhart \& Lazarian 2017, in prep.)\footnote{\url{http://www.mhdturbulence.com}}.

Each simulation is defined by the sonic Mach number $M_s \equiv  |\pmb{\varv}|/c_s$, and the Alfv\'{e}nic Mach number $M_A \equiv |\pmb{\varv}|/ \langle \varv_A \rangle$, where $\pmb{\varv}$ is the velocity, $c_s$  and $\varv_A$ are respectively the isothermal sound speed and the Alfv\'en speed, and $\langle \cdot \rangle$ denotes averages over the entire simulation box.
We explore a range of sonic Mach numbers ($M_S \approx 0.7, 1, 2, 4$, and $7$) for two different regimes of Alfv\'enic Mach number. The simulations are sub-Alfv\'{e}nic with $M_A\approx0.7$ (i.e.~strong magnetic field) or  super-Alfv\'{e}nic (${M}_{A}\approx2.0$). The initial Alfv\'enic Mach number in the super-Alfv\'enic runs is 7.0; however due to strong small-scale dynamo effects \added{(see eg. \cite{federrath_2016})} the magetic field rises for this run and the saturated end value of $M_A$ is roughly 2. The sonic Mach number does not evolve significantly throughout the simulations. The simulations are non-self-gravitating, meaning they are scale-free and we may assign any desired physical scale for the box length and density (see \citealt[][Appendix]{Hill2008}). The sonic Mach number scales as $1/\sqrt{P}$ and the Alfv\'enic Mach number scales as $1/B$; our choices of sonic Mach number correspond to steps in pressure by a factor of 10. Our choice of initial  Alfv\'enic Mach number also corresponds to a step in $B$ by a factor of $10$, but the factor between the two final, saturated magnetic fields is roughly $3$. We show slices of a representative range of these simulations in Figure \ref{fig:sim_boxes}. Throughout, we use $\simeq$ when stating the sonic and Alfv\'enic Mach numbers because they are average quantities over the simulation box and do fluctuate slightly both spatially and in time.

\section{Results}
\label{sec:results}
\begin{figure}
\includegraphics[width=.95\columnwidth]{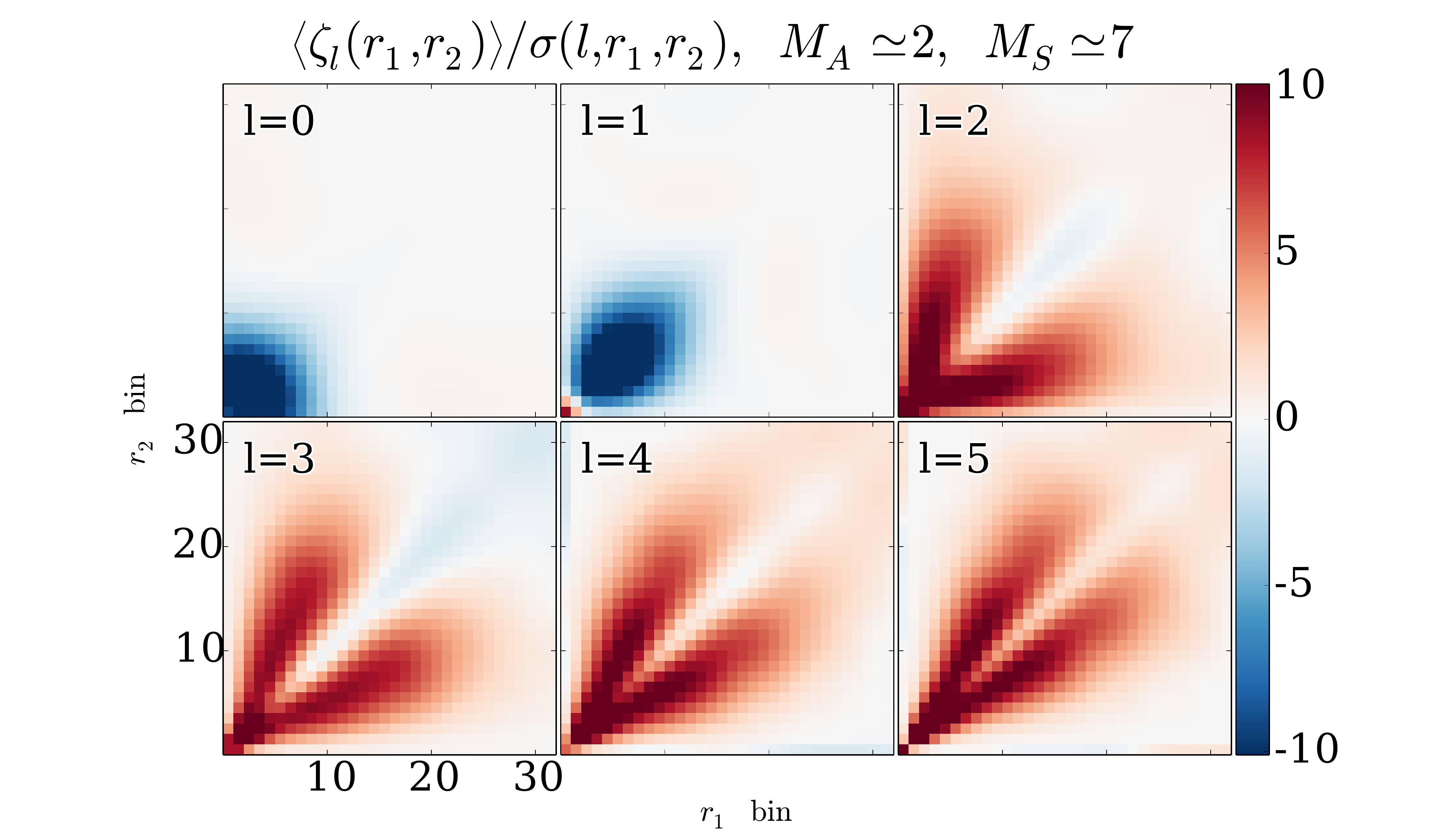}
\includegraphics[width=.95\columnwidth]
{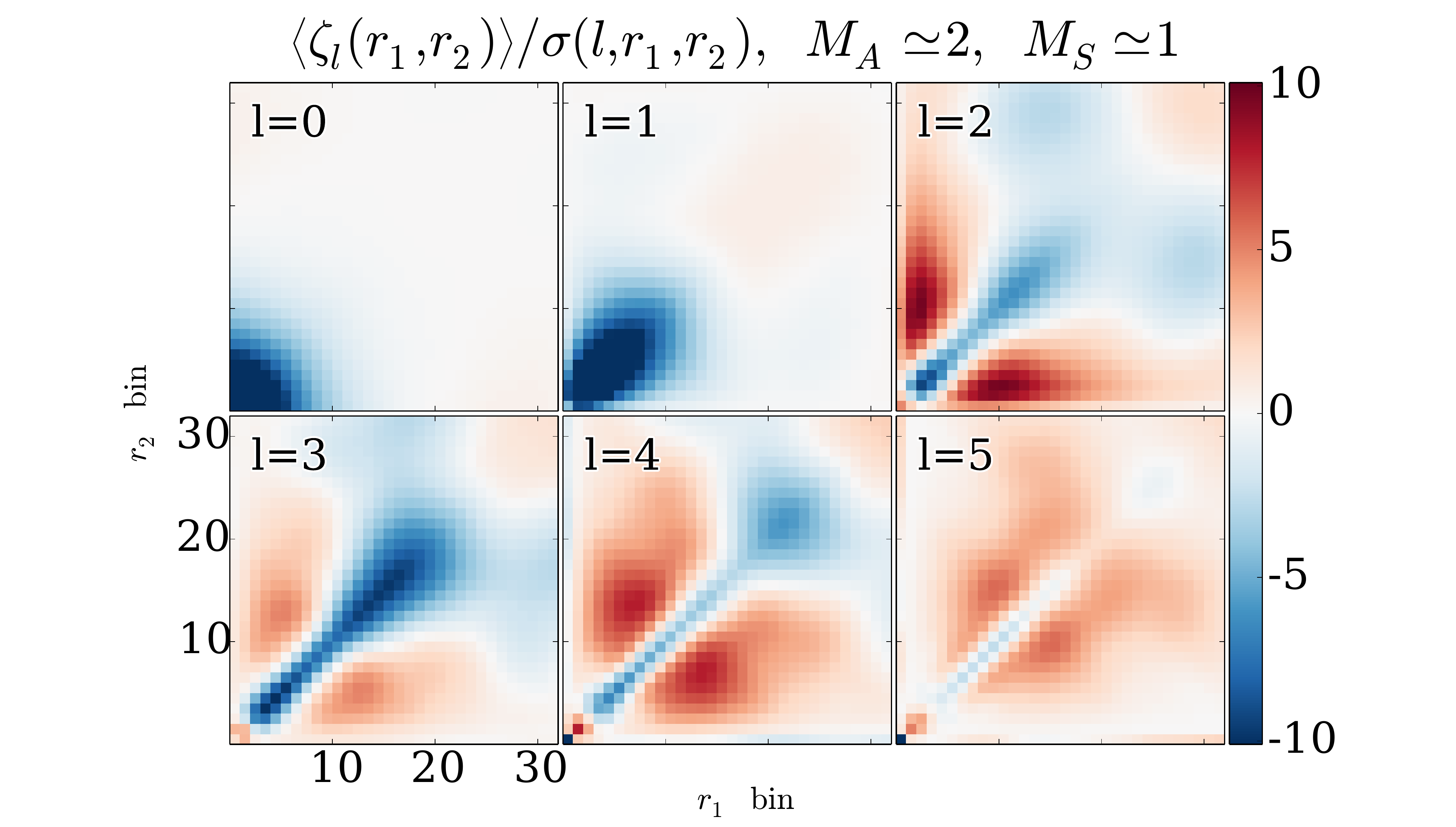}
\includegraphics[width=.95\columnwidth]
{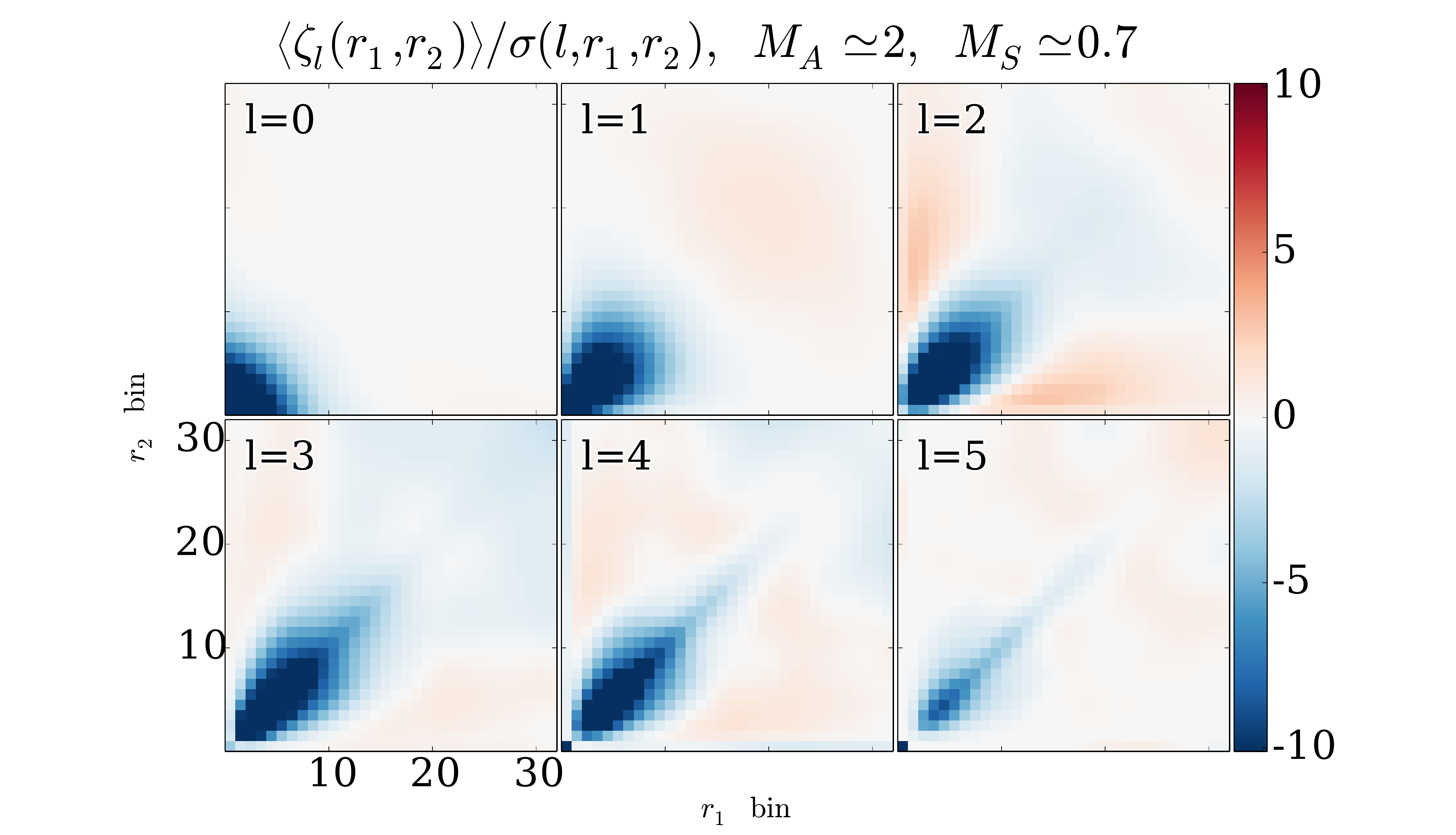}
\caption{The full 3-D 3PCF signal-to-noise computed in the multipole basis, for three different sonic Mach numbers $M_S$, for the super-Alfv\'enic Mach number runs. The top panel is supersonic, the middle panel transonic, and the bottom panel subsonic. Each panel shows the radial coefficients of the 3PCF as projected onto the Legendre basis for the angular dependence; the triangle sides are binned, and denoted $r_1$ and $r_2$. The 3PCFs look very similar in the monopole $\ell=0$, but are easily distinguished by using the angular information contained in the higher multipoles. The supersonic case's density field has the largest number of filaments (see Figure \ref{fig:sim_boxes}), created by shocks. This panel confirms our expectation that filamentary structure will produce a strong diagonal amplitude that grows with $\ell$. This occurs because higher-$\ell$ Legendre polynomials put more weight on the fully closed $\mu = 1$ or fully-opened $\mu = -1$ triangles to which filaments map, as the Legendre polynomials' slopes become steeper with rising $\ell$ as one moves away from $\mu = \pm 1$. The apparent convergence of the higher multipoles $\ell = 3,4$, and $5$ is also consistent with filaments, which in the limit of infinitesimal width have equal power at all multipoles much as a Dirac Delta function is an equal-weight sum of all plane waves.}
\label{fig:MA7_full}
\end{figure}

\begin{figure}
\includegraphics[width=.95\columnwidth]{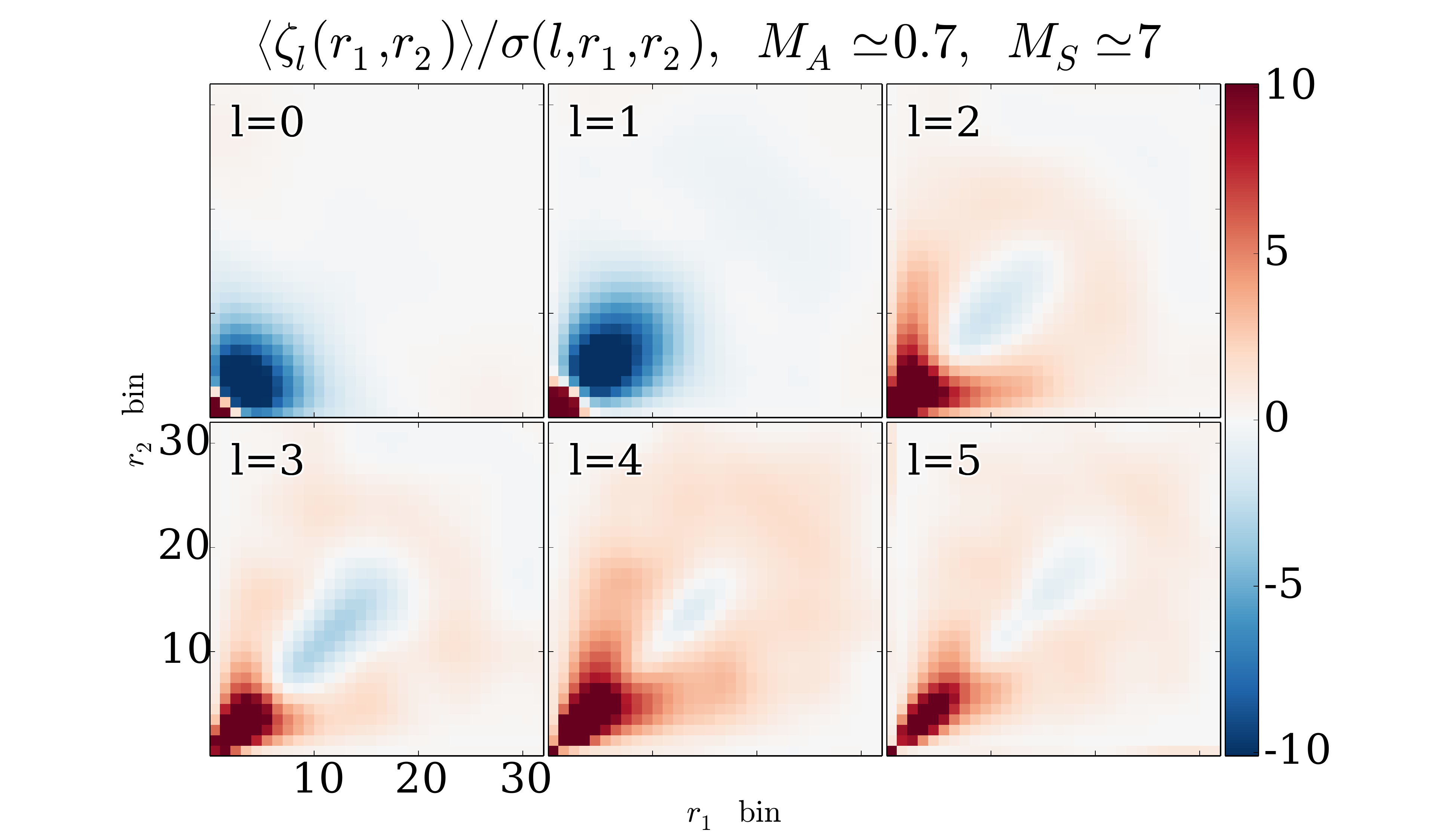}
\includegraphics[width=.95\columnwidth]
{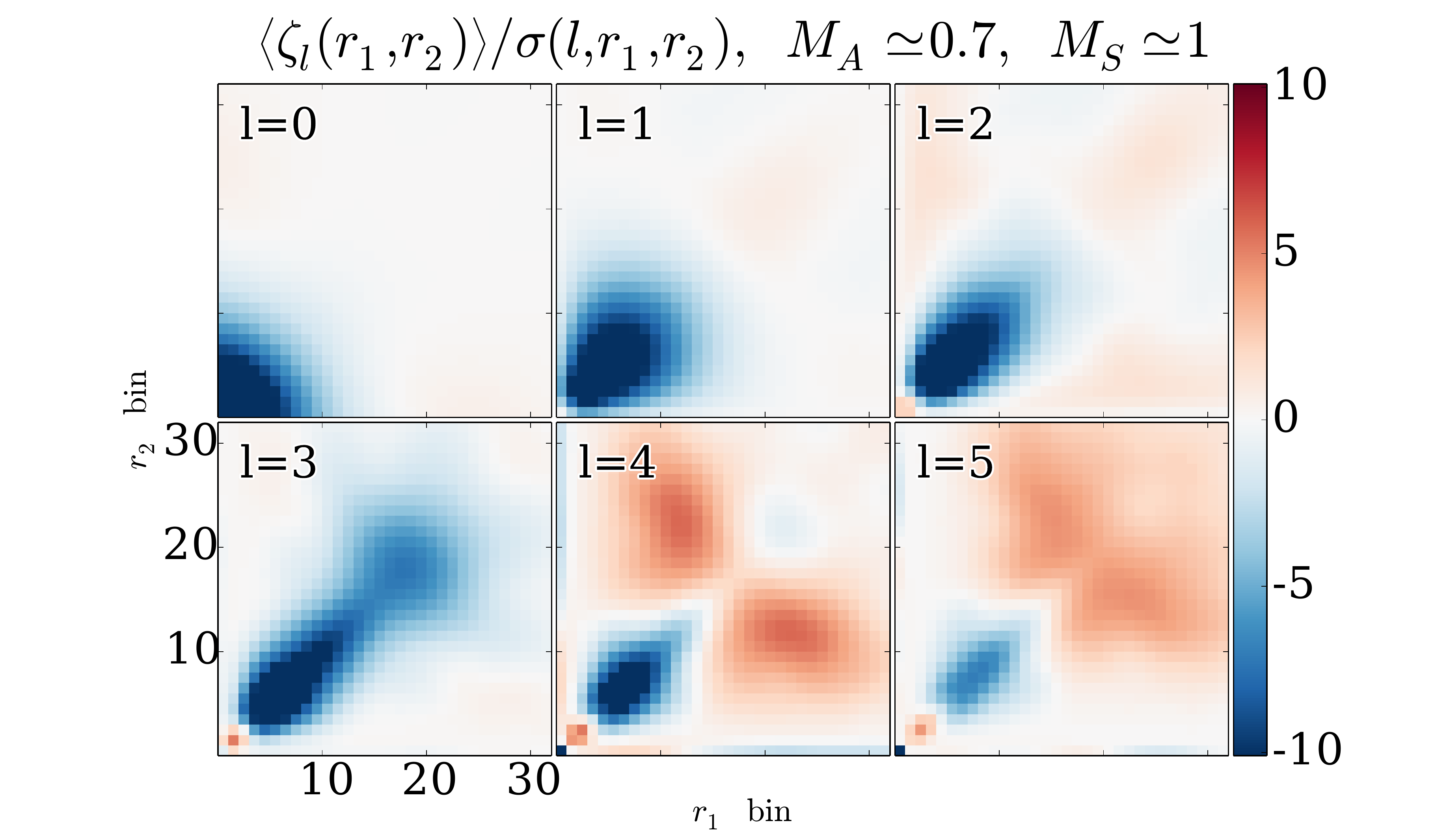}
\includegraphics[width=.95\columnwidth]
{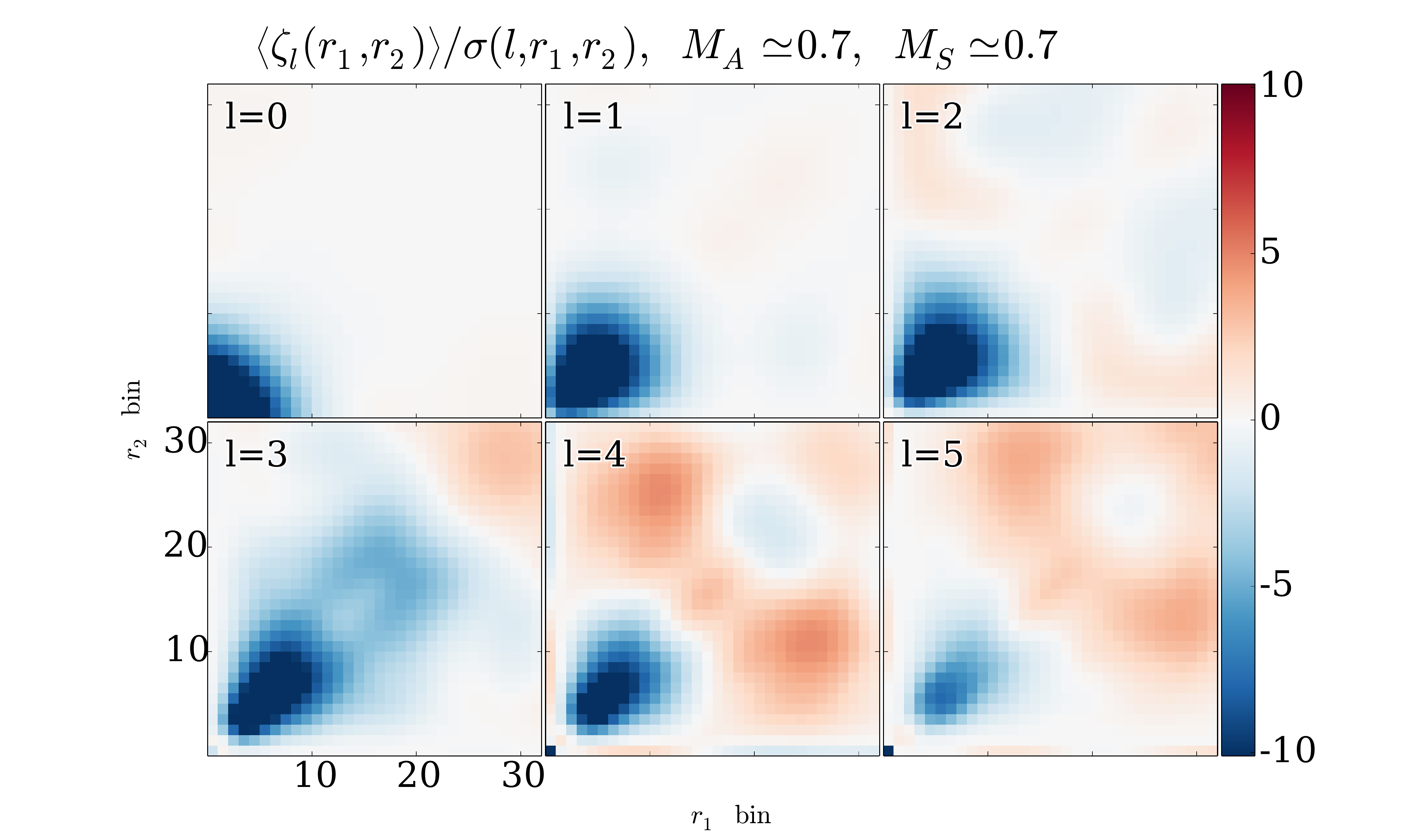}
\caption{Similar to Figure \ref{fig:MA7_full}, except for sub-Alfv\'enic runs. Again, the key point is that the $\ell=0$ moment does not offer much discriminating power among the three conditions, whereas the higher multipoles break the degeneracy. Cross-comparing with Figure \ref{fig:MA7_full} shows that the higher multipoles also enable distinguishing between different Alfv\'en Mach numbers (magnetic fields) at fixed sonic Mach number. Indeed, overall, none of these six plots (Figures \ref{fig:MA7_full} and \ref{fig:MA_pt7_full}) look the same as each other when including the $\ell>0$ multipoles, showing that there is not much degeneracy and in principle the 3PCF can distinguish both different sonic and Alfv\'enic Mach numbers. In practice observational issues such as noise and line-of-sight integration may make this more challenging.}
\label{fig:MA_pt7_full}
\end{figure}

First, we plot the multipoles of the 3PCF as functions of the two triangle side lengths as color maps (Figures \ref{fig:MA7_full} and \ref{fig:MA_pt7_full}). We have chosen for each $M_A$ to display supersonic, transonic, and subsonic cases, respectively $M_S \simeq 7, 1$, and $0.7$. We also studied $M_S \simeq 4$ and $2$ but for the sake of brevity we do not show them; they look similar to the nearest values of $M_S$ (7 and 1) we do show. In these plots, red denotes an excess of triangles above random at some pair of side lengths $r_1,r_2$, projected onto the specified multipole in angle; blue denotes a deficit. Figure \ref{fig:alg_fig} provides a compact guide to interpreting these plots. Our bins are of constant width of 4 simulation pixels each, so going out to 32 bins corresponds to probing scales up to half the box size. It is inappropriate to go to larger scales as in a periodic box, the dynamics already  begins to incur order-unity errors on scales of half the box size. Furthermore, triangles where one side is larger than half the box size will periodically overlap themselves as we take the simulation grid to have periodic boundary conditions, and this effect is unphysical. Bins at larger radii enclose more voxels and thus more possible triangles: we normalize out this effect by dividing the 3PCF by the product of the number of voxels in the two radial bins, which scales as $r_1^2r_2^2$. Furthermore, because the 3PCF multipoles show some time-variation, we average 9 snapshots of the same simulation, separated by half turnover times, and beginning late enough in the simulation that the turbulence is already well-developed. In units of the turnover time, these snapshots are at $5,\;5.5,\;6,\;6.5,\;7,\;7.5\;,8\;,8.5$, and $9$. We then divide each multipole by the standard deviation of its mean to yield the 3PCF multipoles in units of signal-to-noise. The standard deviation is roughly constant as a function of $r_1$ and $r_2$, so dividing by it does not drastically change the scale dependence of the 3PCF multipoles.

We detect significant features in the first few multipoles for all the simulations. In nearly all cases, the $\ell = 5$ features are quite similar to those at $\ell=4$, suggesting that the higher multipoles are converging to a common structure and thus we are not losing much discriminating power by truncating our expansion at $\ell =5$. Generically, across simulations, the monopole ($\ell = 0$) is negative out to bin 10. Similarly, across simulations, the dipole ($\ell = 1$) is negative at small radii and has largest amplitude around the diagonal $r_1=r_2$. In the supersonic and super-Alfv\'enic case, the dipole is positive in the first two bins, and in the supersonic and sub-Alfv\'enic case, both the monopole and dipole are positive for the first few bins. The higher multipoles differ more dramatically across conditions, meaning that they provide important information in distinguishing different conditions.
 
In the highly supersonic cases $M_S\simeq 7$, the high multipoles $\ell \geq 2$ are positive at small radii, then are negative along the diagonal and positive away from the diagonal. In the transonic and subsonic cases $M_S \leq 1$, the behavior of the higher multipoles changes to be negative at small radii. The higher multipoles show significant positive features at larger radii as one moves away from the diagonal.

\begin{figure*}
\includegraphics[scale=.5]{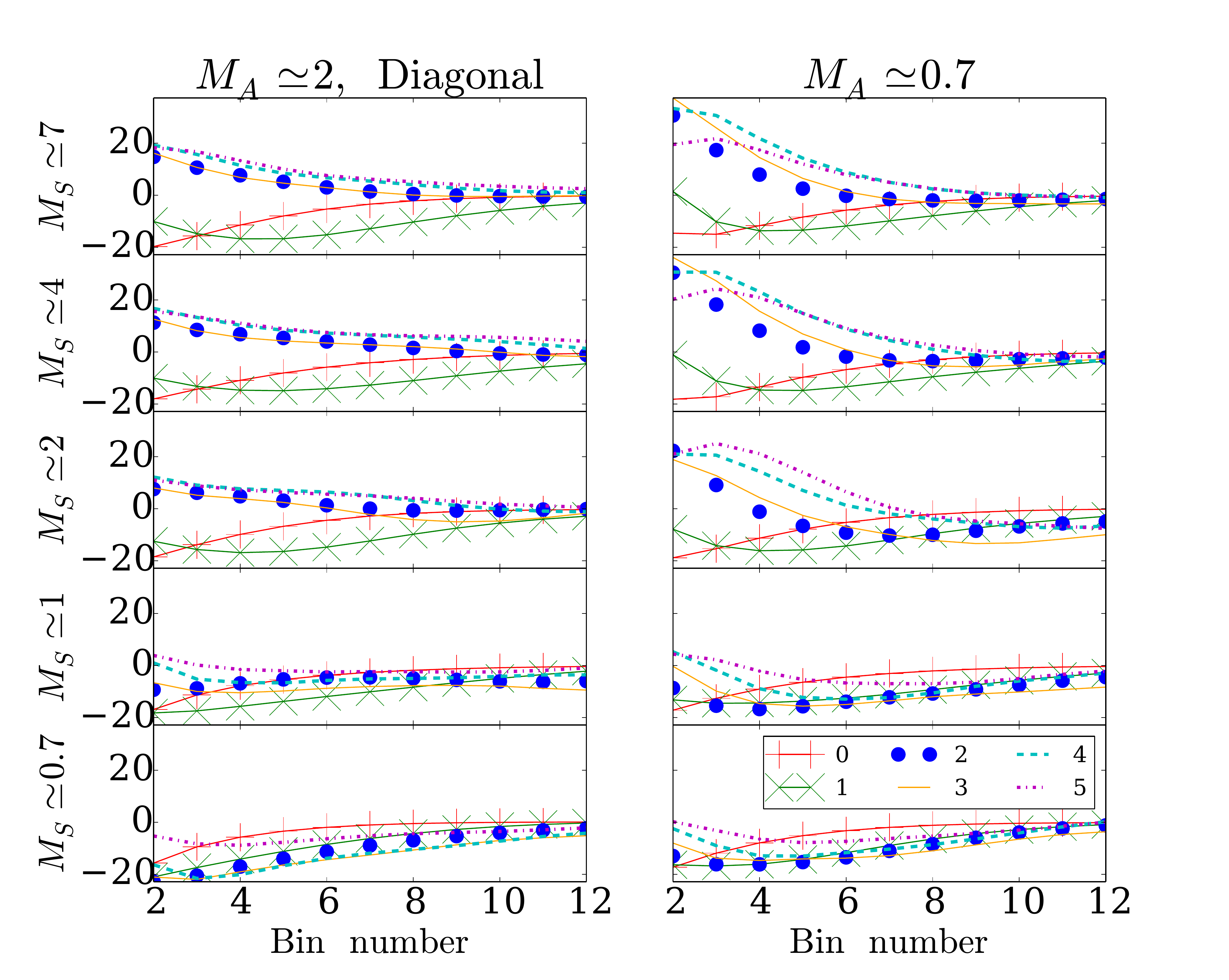}
\caption{Each panel shows the diagonal, $r_1 = r_2$, of the 3PCF for  all $\ell$ at a given sonic and Alfv\'enic Mach number. Visually this is traversing a $45^{\circ}$ line in the panels shown in Figures \ref{fig:MA7_full} and \ref{fig:MA_pt7_full}. Physically it corresponds to isosceles triangles. We show only the first 12 bins (omitting bins 0 and 1 due to likely contamination by numerical noise) here, as the signal becomes smaller beyond bin 12 so the curves become difficult to distinguish by eye. As one moves down the lefthand column above, i.e. varying the sonic Mach number while keeping the Alfv\'enic Mach number (magnetic field) fixed, one sees that the multipole ordering gradually changes; the higher multipoles, $\ell\geq 2$, change from positive (supersonic cases) to negative (transonic and subsonic cases). A similar trend occurs in the righthand column, which has a smaller Alfv\'enic Mach number. The variation is continuous as one moves down each column, showing that the 3PCF has a smooth response to changes in sonic Mach number. Reading across the two columns, at fixed sonic Mach number, it is clearly possible to distinguish between the two Alfv\'enic Mach numbers (supersonic and subsonic) presented here; see also Figures \ref{fig:MA7_full} and \ref{fig:MA_pt7_full}.}
\label{fig:conditions_all_ell}
\end{figure*}

\begin{figure*}
\includegraphics[scale=.5]{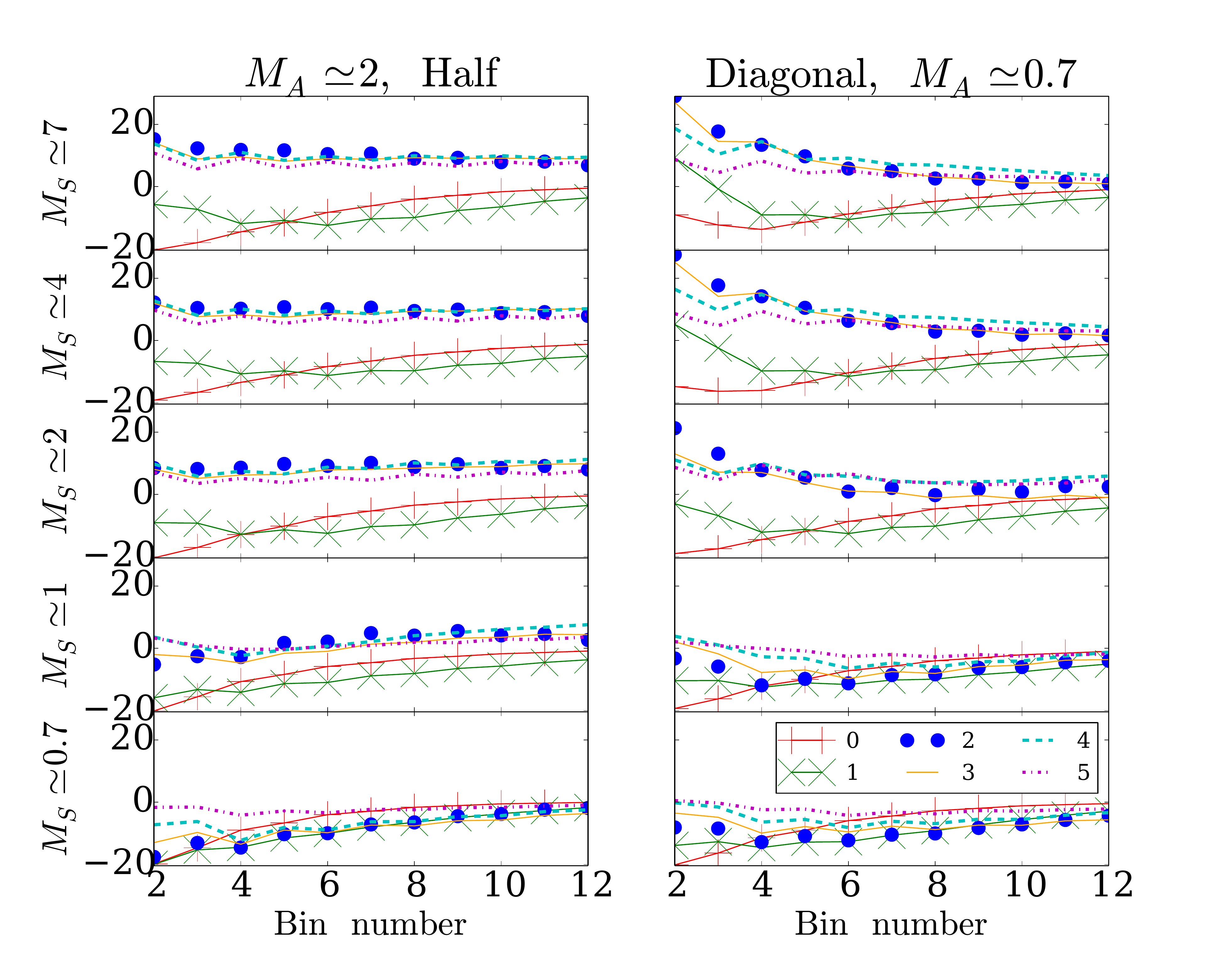}
\caption{Each panel shows the half-diagonal, $r_2 = r_1/2$, of the 3PCF for  all $\ell$ at a given sonic and Alfv\'enic Mach number. Visually this is traversing a $26.6^{\circ}$ line in the panels shown in Figures \ref{fig:MA7_full} and \ref{fig:MA_pt7_full}. Physically it corresponds to $2:1$ triangles. We show only the first 12 bins (omitting bins 0 and 1 due to likely contamination by numerical noise) here, as the signal becomes smaller beyond bin 12 so the curves become difficult to distinguish by eye. As one moves down the lefthand column above, i.e. varying the sonic Mach number while keeping the Alfv\'enic Mach number (magnetic field) fixed, one sees that the multipole ordering gradually changes; the higher multipoles, $\ell\geq 2$, change from positive (supersonic cases) to negative (transonic and subsonic cases). A similar trend occurs in the righthand column, which has a smaller Alfv\'enic Mach number. The variation is continuous as one moves down each column, showing that the 3PCF has a smooth response to changes in sonic Mach number. Reading across the two columns, at fixed sonic Mach number, it is clearly possible to distinguish between the two Alfv\'enic Mach numbers presented here, especially at small scales.}
\label{fig:half_diag_conditions_all_ell}
\end{figure*}

These color plots show our full 3-D 3PCF measurements and highlight important features of the 3PCF, but are difficult to compare in detail by eye. Many important features occur \deleted{on }roughly on the diagonal $r_1 = r_2$ or on the half-diagonal $r_2 = r_1/2$, so we compress the measurement by taking the 3PCF multipole values along these two lines. These diagonal and half-diagonal values for each multipole are now simply a function of radius, enabling their display as line plots.

Figure \ref{fig:conditions_all_ell} shows the diagonals for all multipoles for a single simulation. We show only bins 2 and larger as the first two bins (0 and 1) are likely dominated by numerical noise; the simulation will not accurately resolve structure on the scale of a few pixels. Figure \ref{fig:conditions_all_ell} shows that the sonic Mach number has a large effect: there is a clear divide between the supersonic simulations with $M_S \geq 2$ and the transonic/subsonic simulations with $M_S \leq 1$. This difference is most prominent in the higher multipoles $\ell \geq 2$. In the supersonic cases, the higher multipoles are positive at small radius and fall with radius. But in the transonic/subsonic case, these multipoles are negative at small radius. For all sonic Mach numbers, the lower multipoles $\ell \leq 1$ are negative at small radii and tend towards zero with increasing radius.

While these patterns hold in both the super-Alfv\'enic ($M_A \simeq 2$) and sub-Alfv\'enic ($M_A \simeq 0.7$) simulations, the magnetic field changes the detailed behavior of the multipoles. In the super-Alfv\'enic and supersonic simulations, the higher multipoles follow similar curves, with $\ell=4$ and $\ell = 5$ being more positive than $\ell=2$ and $\ell=3$. But in the sub-Alfv\'enic and supersonic simulations, $\ell=4$ and $\ell=5$ flatten at smaller radii while $\ell=2$ and $\ell=3$ do not.


Figure \ref{fig:half_diag_conditions_all_ell} shows the half-diagonals ($r_2 = r_1/2$) for all multipoles for a single simulation. The half-diagonal will contain information not reflected in the diagonal. In particular, the higher multipoles $\ell \geq 2$ have large contrasts between the diagonal and off-diagonal 3PCF. Again, there is a clear divide between the supersonic simulations with $M_S \geq 2$ and the transonic/subsonic simulations with $M_S \leq 1$, which is most prominent in the higher multipoles $\ell \geq 2$. In the supersonic cases, the higher multipoles are positive and fall slowly with radius. But in the transonic/subsonic case, these multipoles are negative at small radius. For all sonic Mach numbers, the lower multipoles $\ell \leq 1$ are negative at small radii and tend towards zero with increasing radius.

The magnetic field also has a noticeable effect on the half-diagonal. In the supersonic simulations, the half-diagonal values for all multipoles are more positive at small radii in the sub-Alfv\'enic simulations than in the corresponding super-Alfv\'enic simulations. In the super-Alfv\'enic simulations that are also subsonic or transonic, the higher multipoles $\ell \geq 2$ do not intersect the lower multipoles $\ell \leq 1$. But in the sub-Alfv\'enic simulations that are subsonic or transonic, the higher multipoles intersect the lower multipoles around bin 6.

\begin{figure*}
\includegraphics[scale=.5]{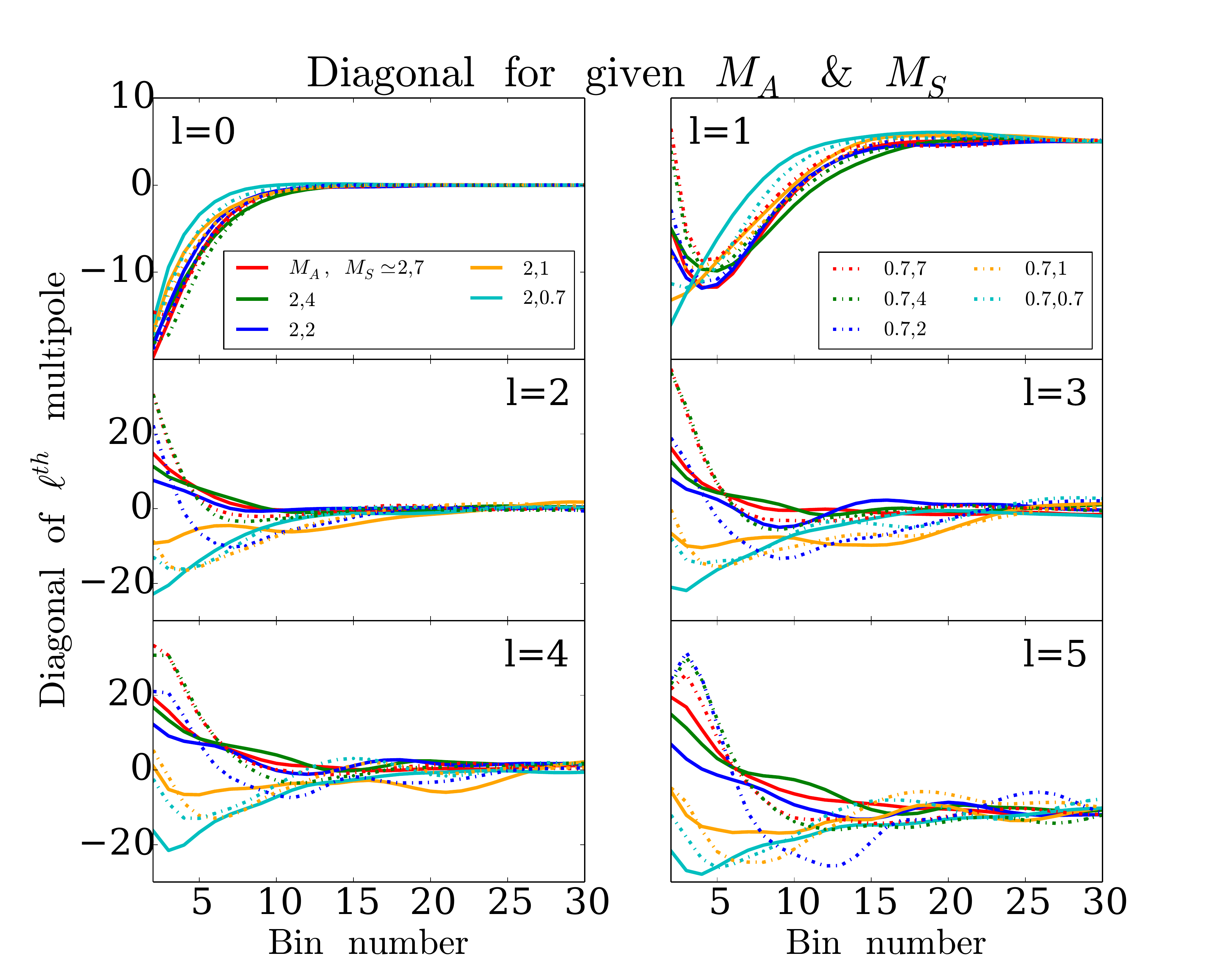}
\caption{Here we show the 3PCF diagonal from bins 2 to 30 for each $\ell$ and all conditions. The information is the same as in Figure \ref{fig:conditions_all_ell} (though we go out to a larger maximal bin here), but the slicing is different. The super-Alfv\'enic cases are displayed with solid lines, sub-Alfv\'enic with dashed lines.  Here the key point is that the $\ell=0$ diagonal does not separate the different conditions well, whereas adding higher multipoles clearly does. In particular, on smaller scales there are significant differences between the cases' diagonals at fixed $\ell$.}
\label{fig:diagonal_each_ell}
\end{figure*}

We highlight the discriminatory power of the diagonal of each multipole in Figure \ref{fig:diagonal_each_ell}. Each panel shows the diagonal values for a single multipole for all simulations. Again, we see the strong effect of sonic Mach number: at fixed $M_A$, the $M_S \geq 4$ cases cluster together and so do the $M_S \leq 1$ cases. The $M_S \simeq 2$ case is between these two clusters, but more often closer to the $M_S \geq 4$ cluster. While these differences due to sonic Mach number manifest in the lower multipoles $\ell \leq 1$, they are more prominent in the higher multipoles  $\ell \geq 2$. The effect of the magnetic field is clearest at small radii in the higher multipoles: the supersonic $M_S \geq 2$ lines are more positive in the sub-Alfv\'enic case than in the super-Alfv\'enic case (save for $M_S \simeq 1$).

\begin{figure*}
\includegraphics[scale=.5]{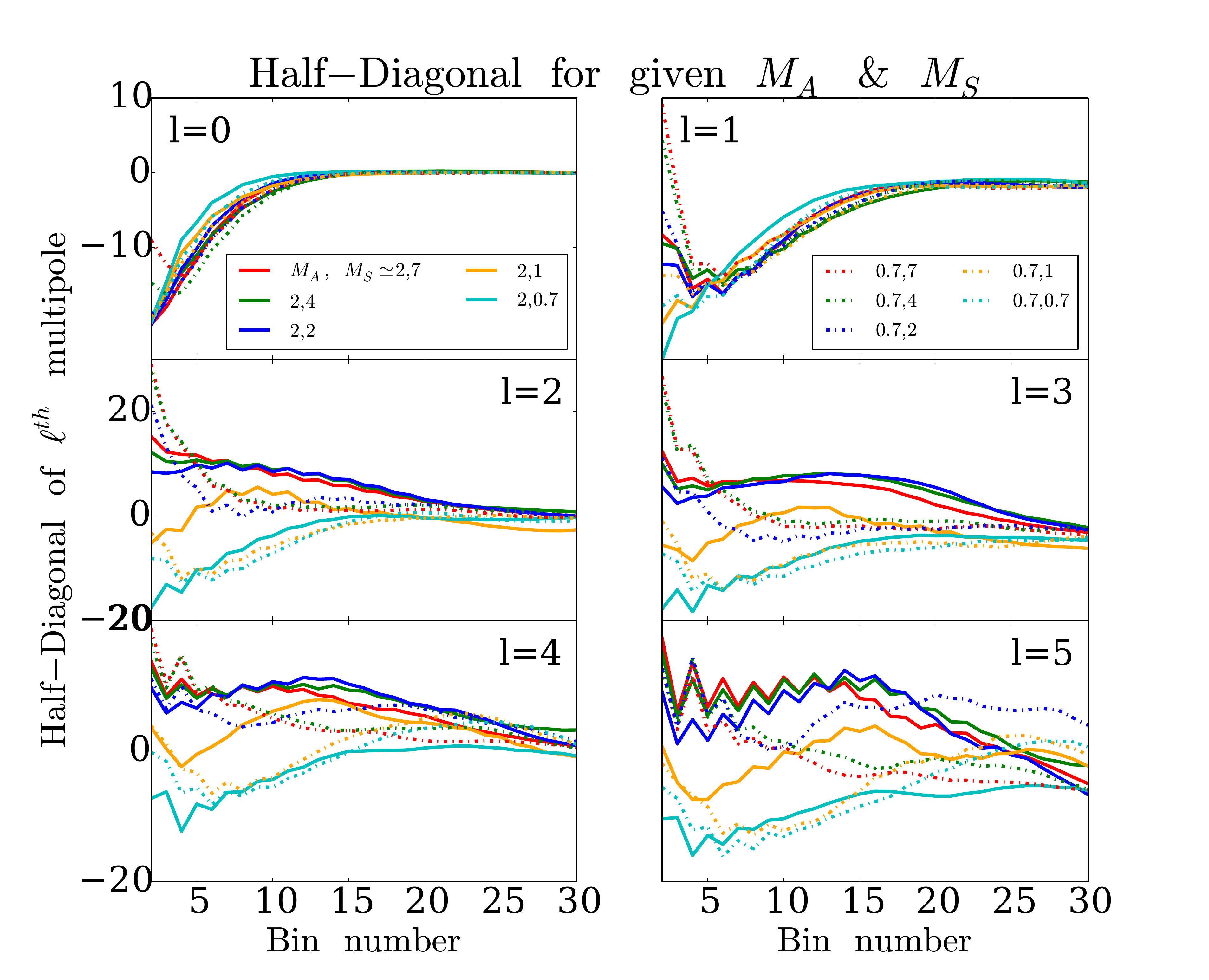}
\caption{Here we show the 3PCF half-diagonal from bins 2 to 30 for each $\ell$ and all conditions. The information is the same as in Figure \ref{fig:half_diag_conditions_all_ell} (though we go out to a larger maximal bin here), but the slicing is different. The super-Alfv\'enic cases are labeled with solid lines, sub-Alfv\'enic in dashed lines.  Again, the $\ell=0$ half-diagonal does not separate the different conditions well, whereas adding higher multipoles clearly does. In particular, on smaller scales there are significant differences between the cases' diagonals at fixed $\ell$.}
\label{fig:half_diag_each_ell}
\end{figure*}

Each panel of Figure \ref{fig:half_diag_each_ell} shows the half-diagonal values for a single multipole for all simulations. In all multipoles, at fixed $M_A$, the $M_S \geq 2$ cases cluster together. The $M_S \simeq 1$ case is always between the $M_S \simeq 0.7$ and $M_S \simeq 2$ cases, often being closer to the $M_S \simeq 0.7$ case. These differences due to sonic Mach number are most prominent in the higher multipoles $\ell \geq 2$ but can also be seen in the dipole $\ell = 1$. The magnetic field also affects the half-diagonal. This effect can be clearly seen in $\ell = 1,2,3$ at small radii: the values are more positive in the sub-Alfv\'enic case than in the super-Alfv\'enic case.


\begin{figure}
\includegraphics[width=1.10\columnwidth]{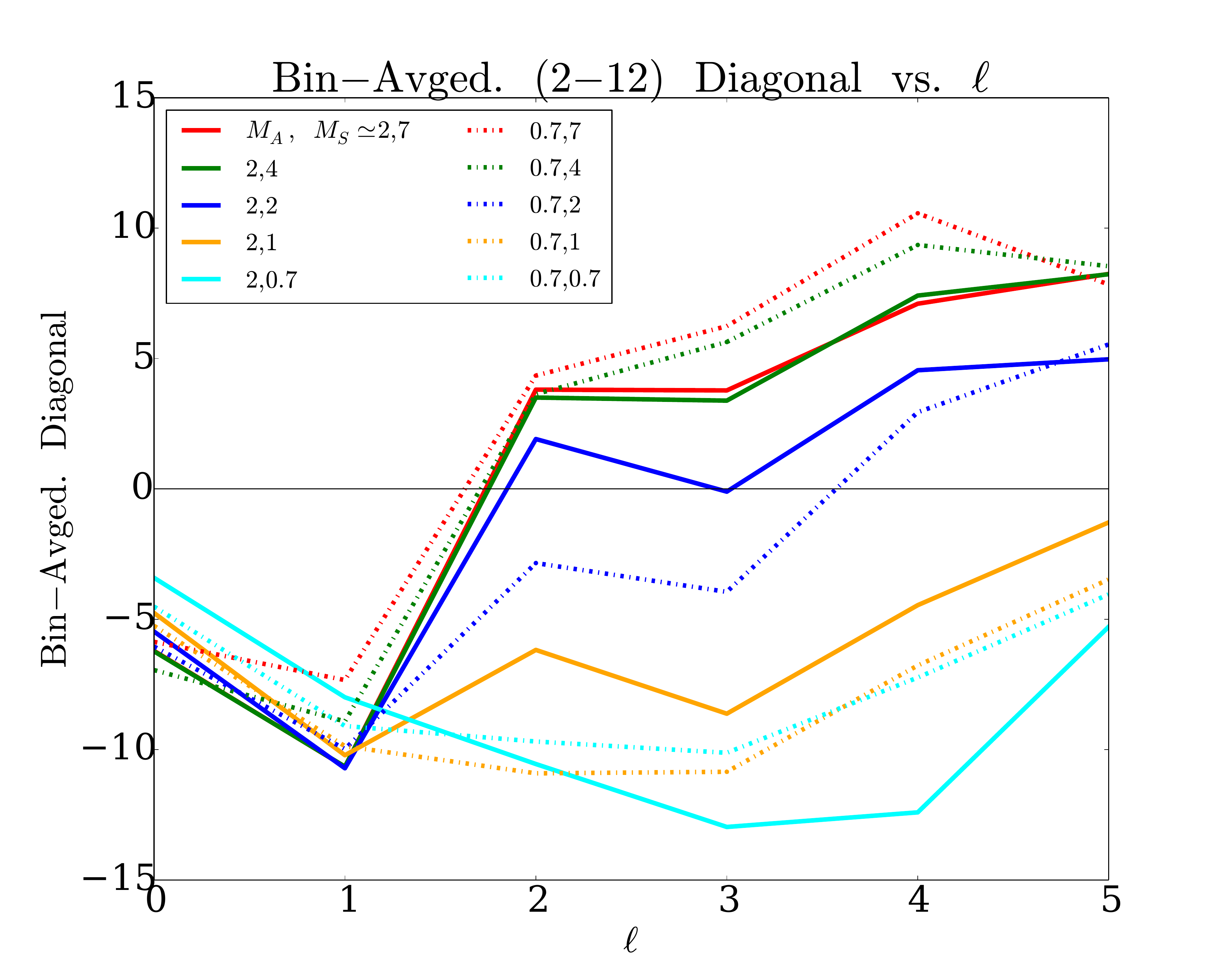}
\includegraphics[width=1.10\columnwidth]{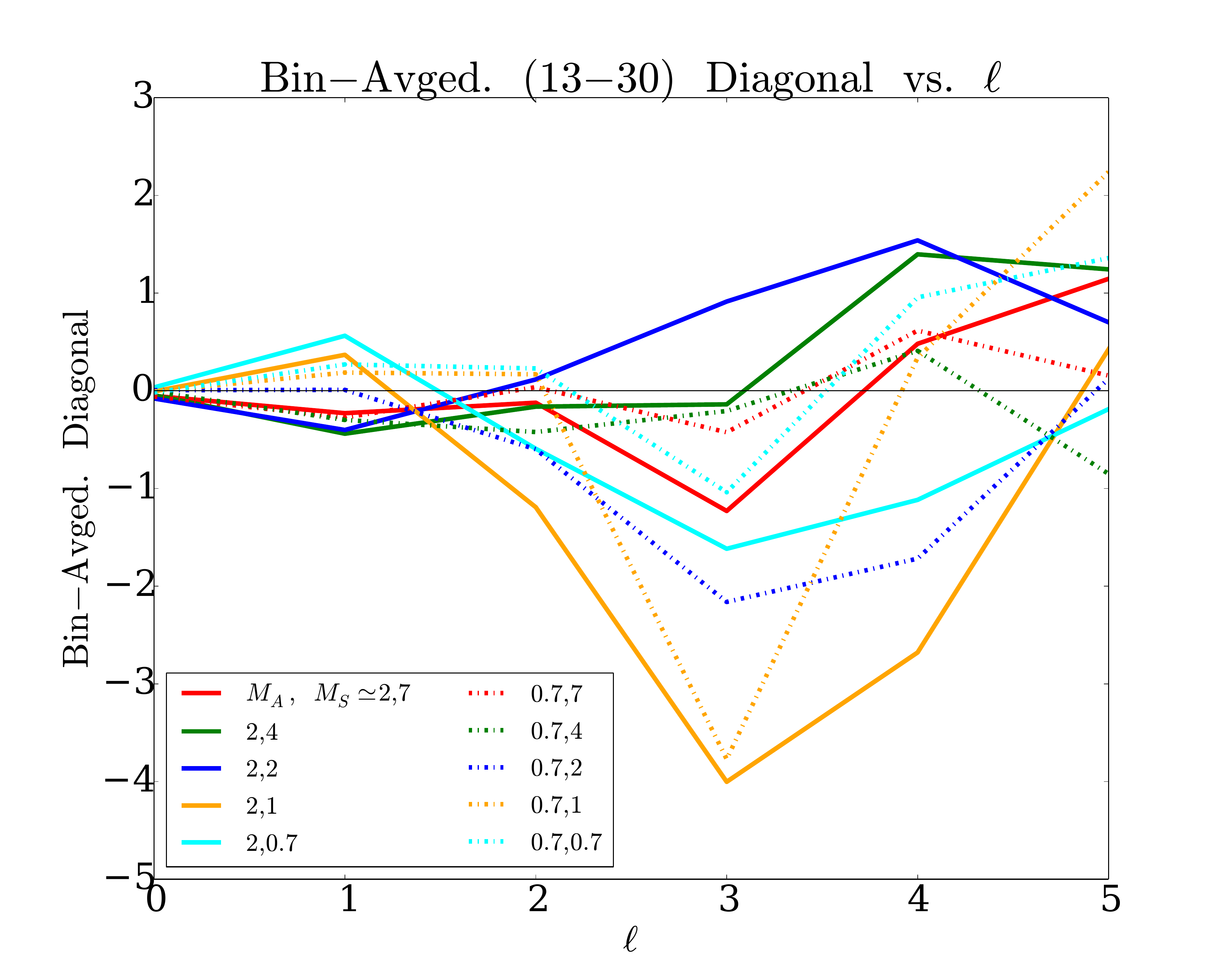}
\caption{The average along the diagonal over the range of bins (inclusive) indicated in the panel title as a function of $\ell$ for all conditions. Visually these panels are the sum over bins of Figure \ref{fig:diagonal_each_ell}, split into two ranges, 2-12 and 13-30. The solid lines are super-Alfv\'enic, the dashed lines sub-Alfv\'enic.}
\label{fig:small_avg}
\end{figure}

\begin{figure}
\includegraphics[width=1.10\columnwidth]{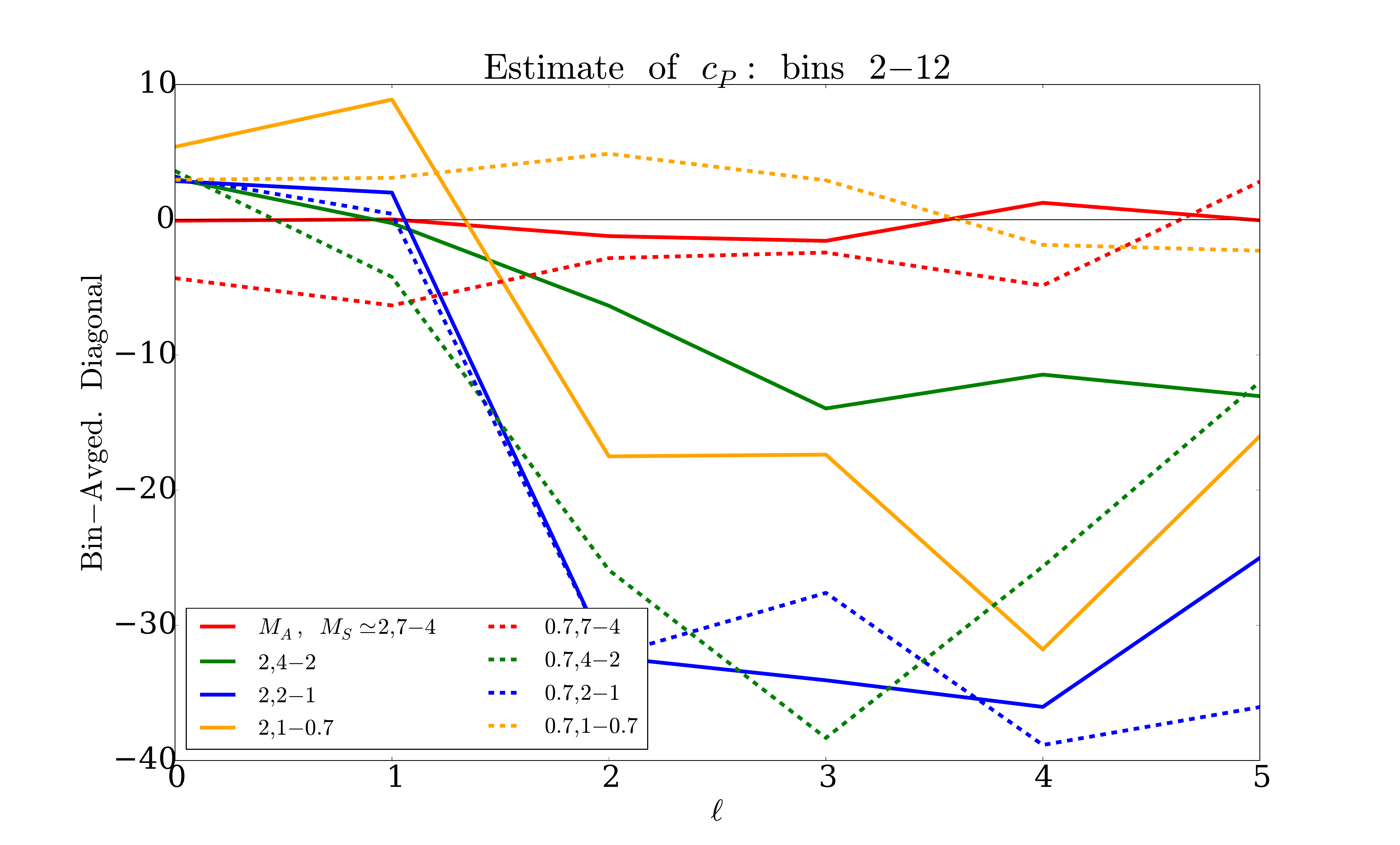}
\includegraphics[width=1.10\columnwidth]{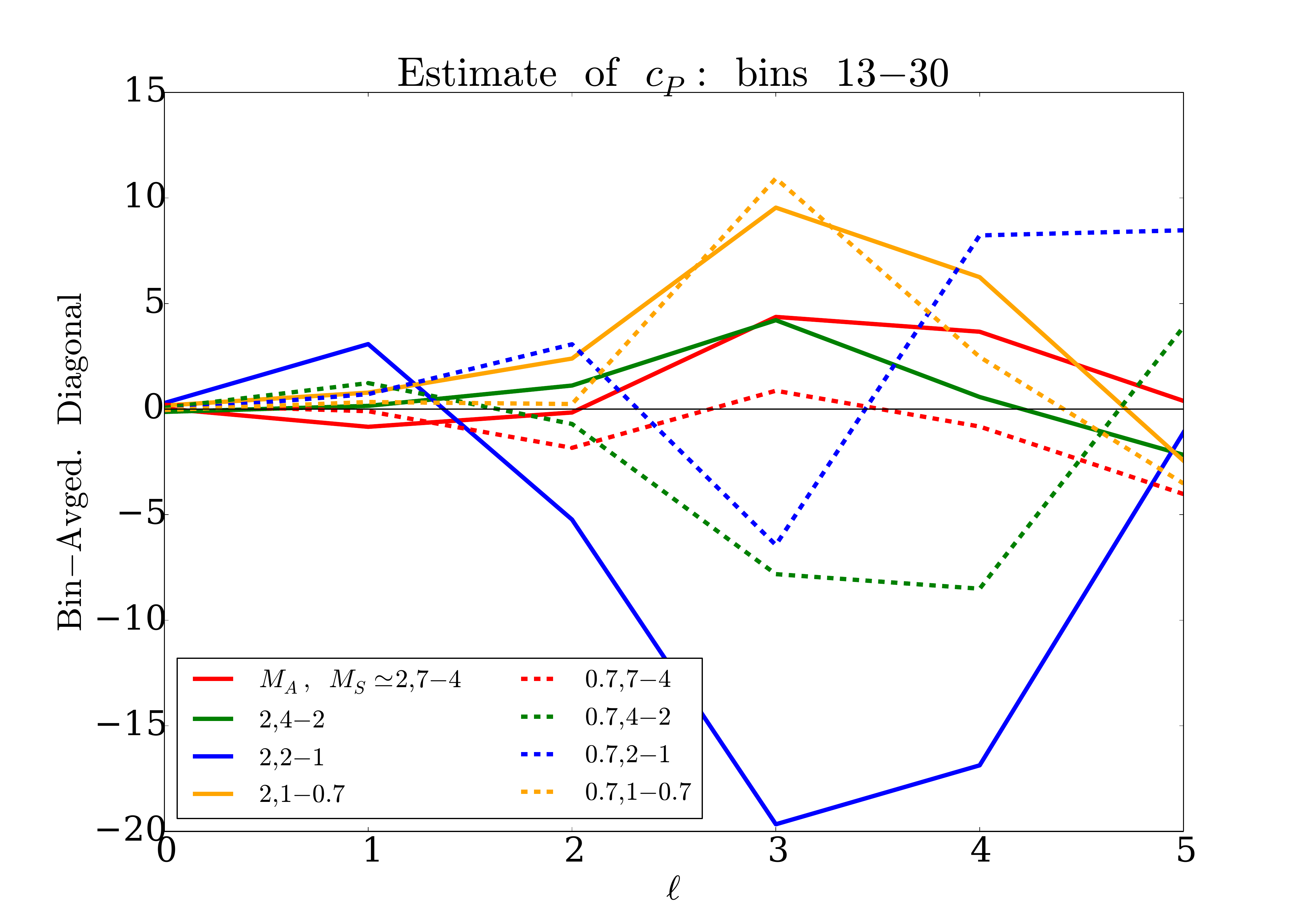}
\caption{$c_P$, the derivative $d \zeta/d[\log_{10} M_S]$, estimated from the two different bin ranges shown in the panel titles as described in \S\ref{sec:results}; see also equation (\ref{eqn:taylor}). The solid lines are $c_P$ estimated from different $M_S$ but the same super-Alfv\'enic $M_A \simeq 2$, the dashed lines $c_P$ estimated from different $M_S$ but the same sub-Alfv\'enic $M_A \simeq 0.7$. This point is also made by the smallness of the $c_B$ (derivatives with respect to $M_A$) compared to the $c_P$ (derivatives with respect to $M_S$). Comparing this Figure with Figure \ref{fig:cB_small}, it is evident the 3PCF is simply less sensitive to $M_A$ than to $M_S$. Were our ansatz for the 3PCF equation (\ref{eqn:taylor}) accurate, $c_p$ would be a constant and all of the curves shown on these panels would overlap. In contrast, we see that $c_p$ is not a constant but rather depends on the pair of $M_S$ used to estimate it, although there is a narrower spread in the monopole $(\ell = 0)$ and dipole $(\ell =1)$. These panels show that the 3PCF's sensitivity to $M_S$, especially of its higher multipoles, is more than a simple power law.}
\label{fig:cp_small}
\end{figure}

\begin{figure}
\includegraphics[width=1.10\columnwidth]{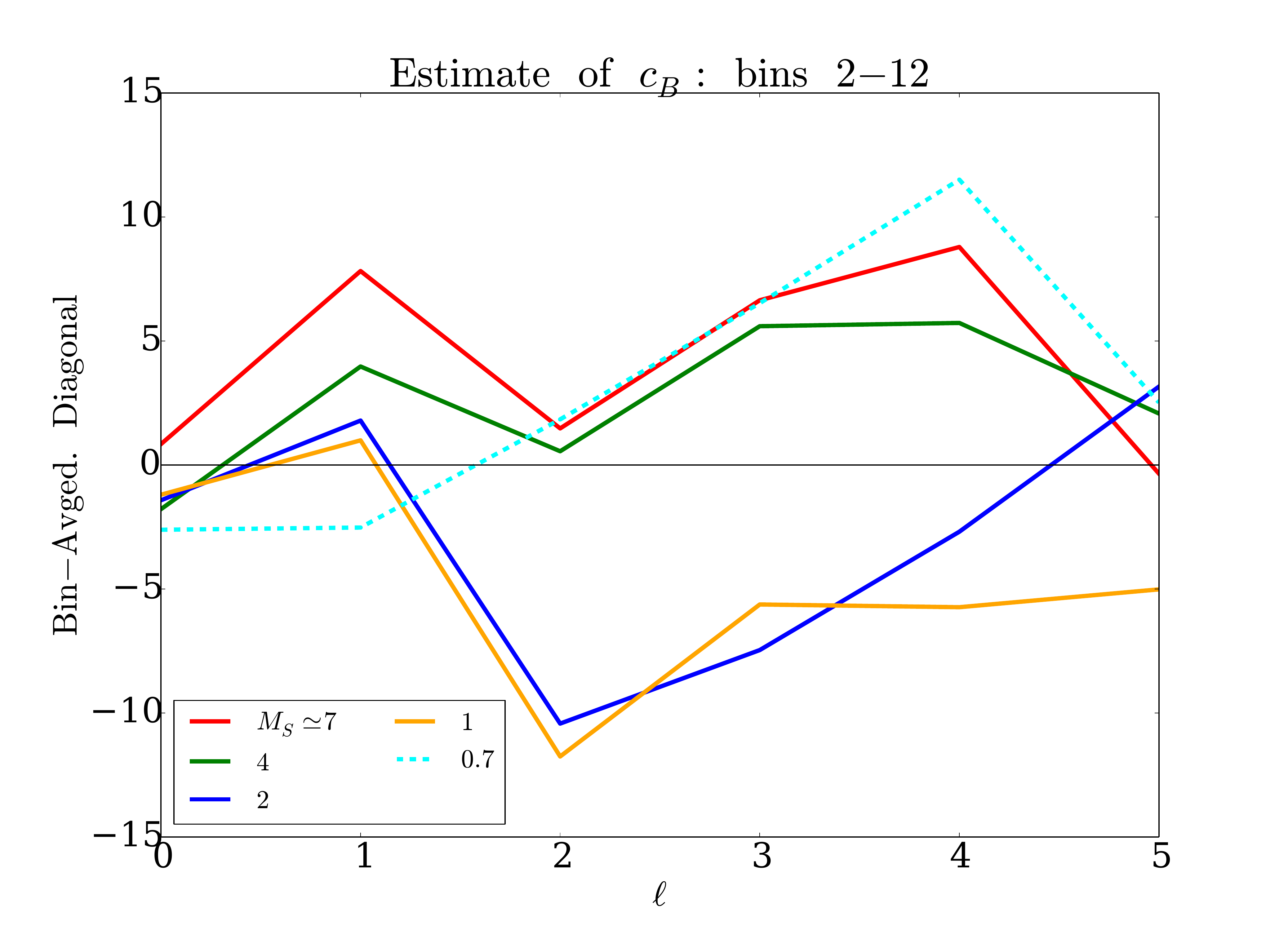}
\includegraphics[width=1.10\columnwidth]{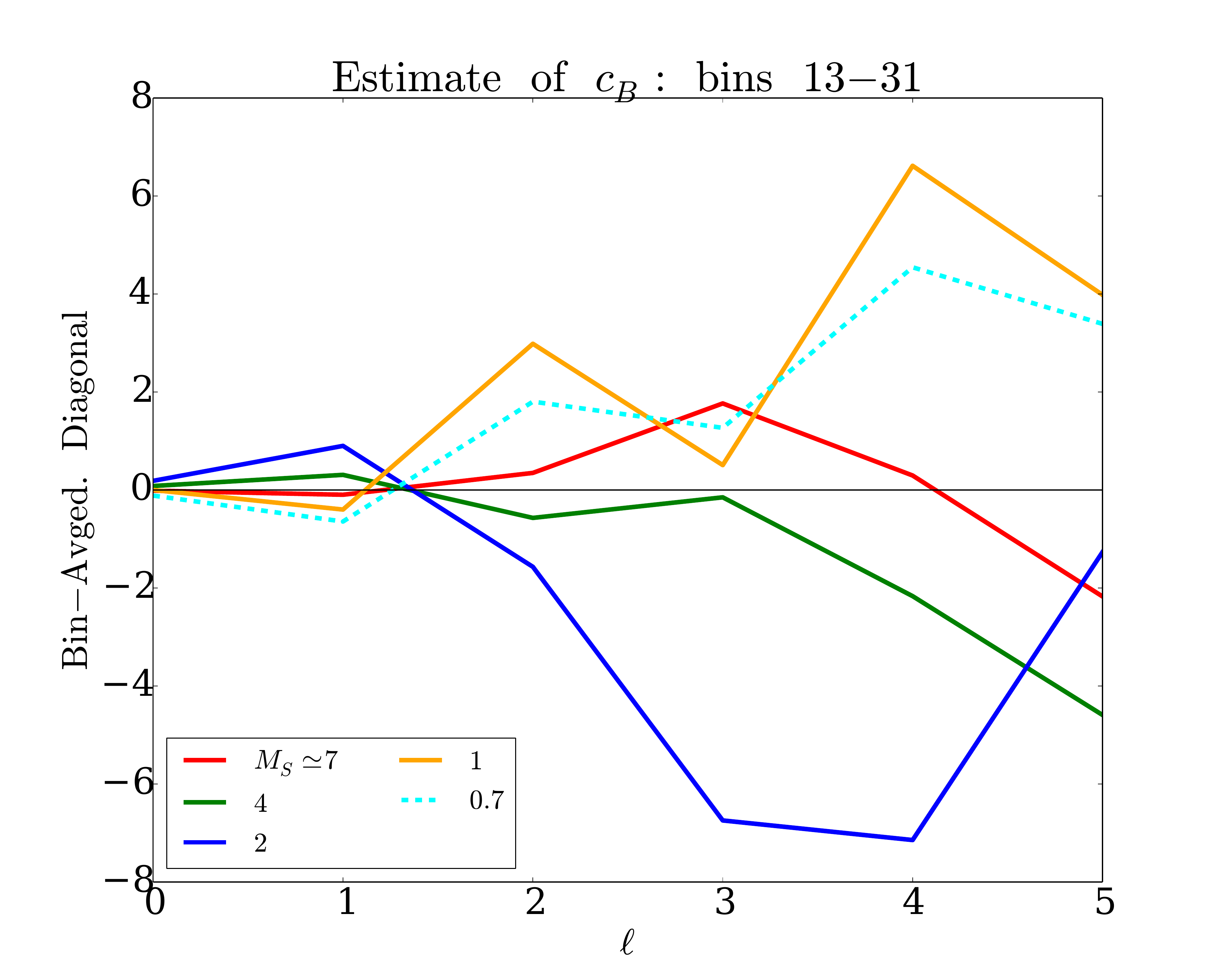}
\caption{$c_B$, the derivative $d \zeta/d[\log_{10} M_A]$, estimated from the two different bin ranges shown in the panel titles as described in \S\ref{sec:results}. Were our ansatz for the 3PCF equation (\ref{eqn:taylor}) accurate, $c_B$ would be a constant and all of the curves shown on these panels would overlap. In contrast, we see that $c_B$ is not a constant but rather depends on the value of $M_S$ at which we differenced the two $M_A$ values used for all the $d \log_{10} M_A$ derivatives. There is a narrower spread in the monopole $(\ell = 0)$ and dipole $(\ell =1)$. But overall these panels show that 3PCF's sensitivity, especially of the higher multipoles, to $M_A$ is more than a simple power law, and that the effect of $M_A$ in fact depends on the value of $M_S$ present.}
\label{fig:cB_small}
\end{figure}

\begin{figure}
\includegraphics[width=1.10\columnwidth]{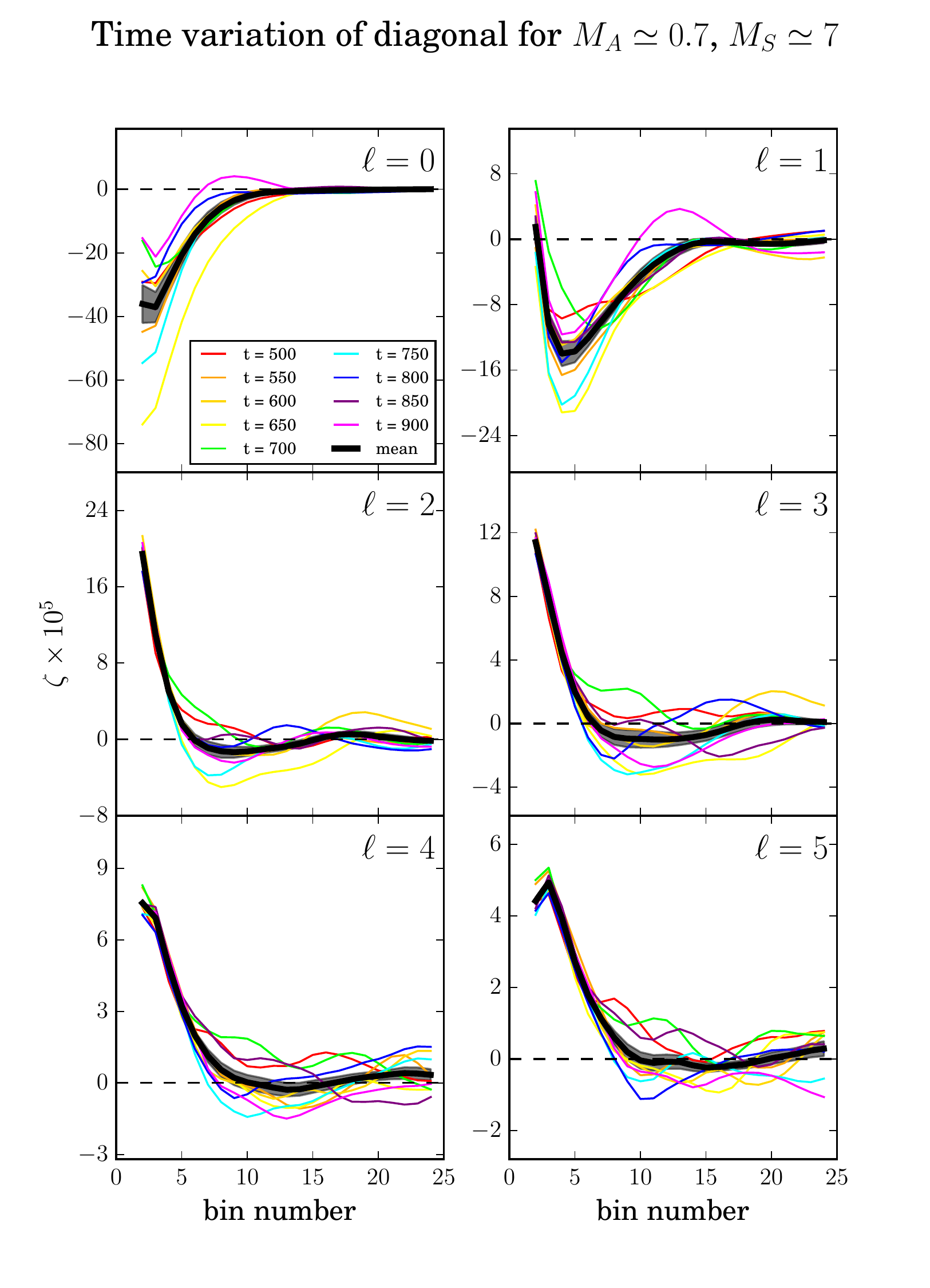}
\caption{These panels show the diagonals of the 3PCF multipoles from bin 0 to 25 at each timestep, as well as the mean (dark, thicker curve) and a range one standard deviation of the mean around the mean (grey shaded band). Here we show a sub-Alfv\'enic and supersonic case; the eddy turnover time is $100$ in our units, so the time-steps are taken every half-eddy-turnover time, starting at $500$ so that the turbulence is well-developed. These illustrate that while there is considerable fluctuation in the 3PCF from time step to timestep, the mean is fairly stable in time, as shown by the tightness of the grey $1\sigma$ band around it. }
\label{fig:saxplot_supersonic}
\end{figure}

\begin{figure}
\includegraphics[width=1.10\columnwidth]{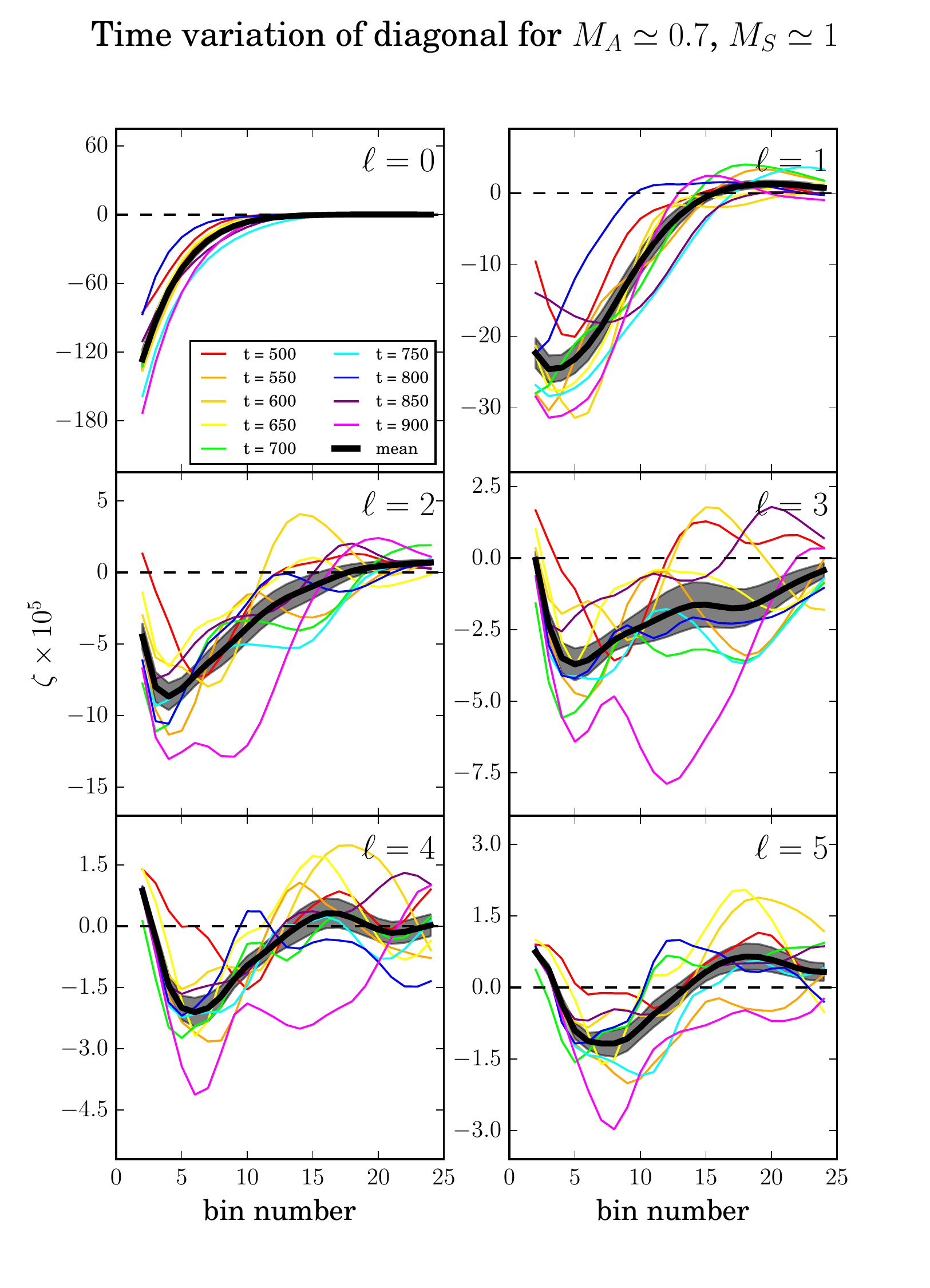}
\caption{Same as Figure \ref{fig:saxplot_supersonic} but for a sub-Alfv\'enic and transonic case. Again the time-stability of the mean is good.}
\label{fig:saxplot_transsonic}
\end{figure}

\begin{figure}
\includegraphics[width=1.10\columnwidth]{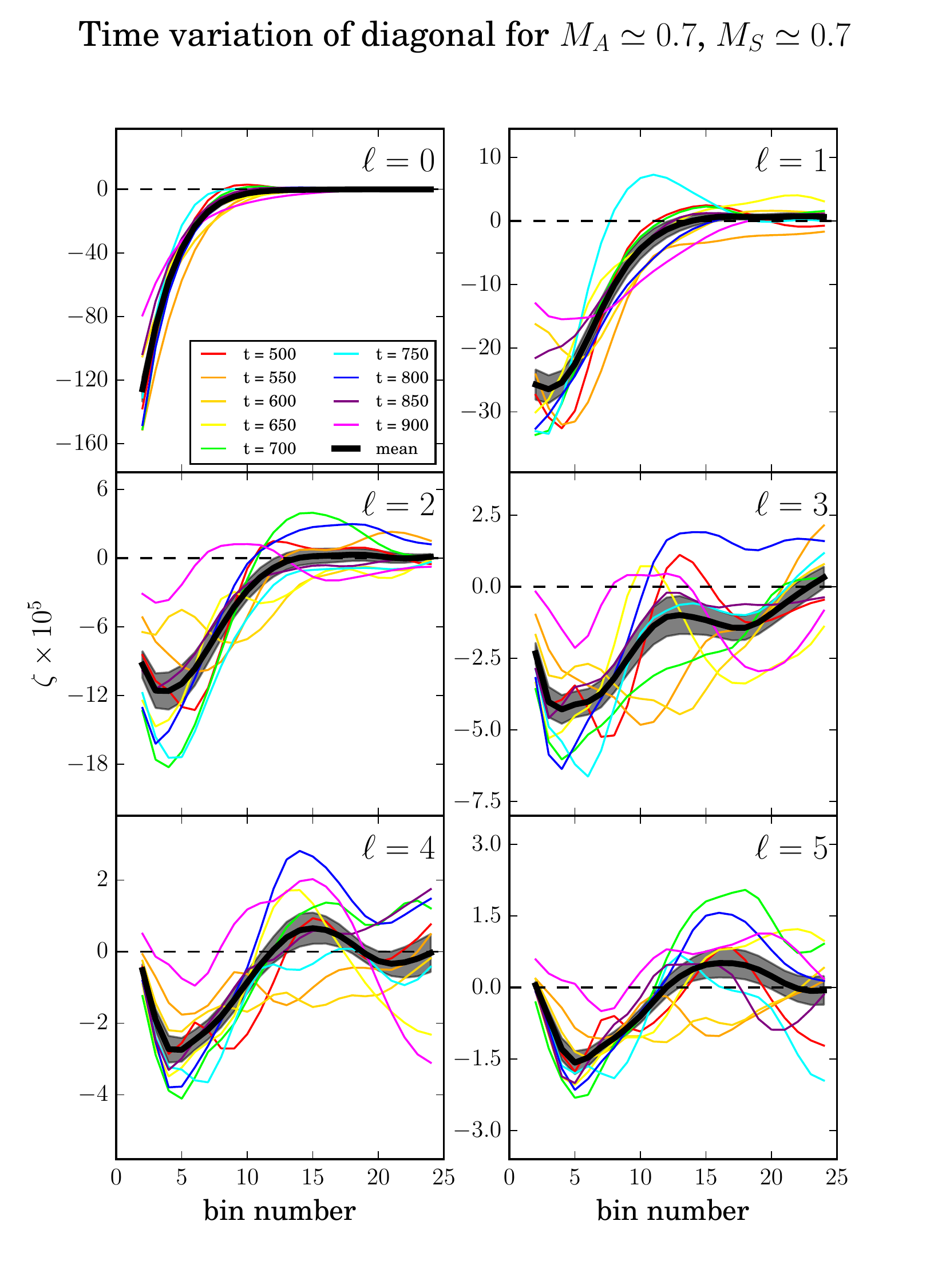}
\caption{Same as Figure \ref{fig:saxplot_supersonic} but for a sub-Alfv\'enic and subsonic case. Again the time-stability of the mean is good.}
\label{fig:saxplot_subsonic}
\end{figure}

We further compress the 3PCF multipoles by averaging the diagonals over two radius ranges, bins 2 to 12 and bins 13 to 30, both inclusive. This compression enables displaying the averaged diagonal for each simulation as a function of $\ell$. The signal is higher for the smaller bin range but the larger range is more comparable to the ``inertial range'' typically found to follow a power-law in the Fourier-space power spectrum of turbulence.\footnote{Though the correspondence is not exact because each configuration-space result is an integral over all Fourier-space modes.} In Figure \ref{fig:small_avg} the effect of sonic Mach number can be clearly seen for bins 2 to 12. The higher multipoles $\ell \geq 2$ are most positive for the supersonic $M_S \geq 4$ cases and lowest for the transonic/subsonic $M_S \leq 1$ cases, with the $M_S \simeq 2$ cases being intermediate. Moving from sub-Alfv\'enic to super-Alfv\'enic renders these higher multipoles more positive. In Figure \ref{fig:small_avg}, the effects of sonic Mach number and Alfv\'enic Mach number are less clear for bins 13 to 30. At these larger radii, features are not very significant: few points are beyond a mean signal-to-noise of 2.

Using simulations at the same Alfv\'enic Mach number and neighboring sonic Mach numbers, we estimate the logarithmic derivatives of the bin-averaged diagonal with respect to $M_S$ and present them in Figure \ref{fig:cp_small}; this is the coefficient $c_P$ defined in equation (\ref{eqn:taylor}). $M_S$ scales as $P^{1/2}$ so $d\log_{10} M_S = -(1/2)d\log_{10} P$. Our simulations with neighboring $M_S$ differ in pressure by a factor of $10^{1/2}$, so $d\log_{10} M_S = 1/4$ (taking the larger $M_S$ minus the smaller $M_S$). We can thus estimate the base-10-logarithmic derivatives with respect to $M_S$ as finite differences by subtracting neighboring pairs of simulations and multiplying by $4$. Our ansatz that the logarithmic derivative is constant at a given multipole does not hold. For $M_A \simeq 0.7$, the pressure derivative is smallest in magnitude for the two most extreme cases, between $M_S \simeq 7,4$ and $M_S \simeq 1,0.7$, and largest in magnitude for the intermediate cases, between $M_S\simeq 4,2$ and $M_S \simeq 2,1$. However, for $M_A \simeq  2$, the pressure derivative is most negative when comparing the estimates of it from $M_S \simeq 2,1$ and $M_S \simeq 1,0.7$.


Comparing simulations at with the same $M_S$ but with $M_A$ of 2 and 0.7, we can also estimate the base-10 logarithmic derivative of the bin-averaged diagonal with respect to $M_A$ using a finite difference; this is the coefficient $c_B$ defined in equation (\ref{eqn:taylor}). $M_A$ varies by a factor of $2/0.7 =2.857$, leading to  $d\log_{10} M_A = 0.456$, meaning we can subtract simulations with neighboring $M_A$ and multiply by $1/0.456 = 2.193$.  The resulting logarithmic derivatives with respect to $M_A$ are shown in Figure \ref{fig:cB_small}. Again, our ansatz that the logarithmic derivative is constant with pressure does not hold. These derivatives are smaller than those for pressure, showing that the 3PCF is more sensitive to pressure than to magnetic field. The supersonic cases $M_S \geq 4$ have similarly structured derivatives: their derivatives are positive for all multipoles and are strongest for $\ell=1,4$. The transsonic/subsonic cases $M_S \leq 1$ are most negative for $\ell = 2$.

We also use the diagonal values of the 3PCF multipoles to evaluate the time variation in the 3PCF. Here we do not divide the 3PCF by the standard deviation of the mean. We show subsonic, transonic, and supersonic at fixed (sub-Alfv\'enic) magnetic field. We do not show super-Alfv\'enic as the result is similar, consistent with the idea that the 3PCF is less sensitive to $M_A$ than to $M_S$, shown by the smaller magnitudes of $c_B$ relative to those of $c_P$; compare Figures \ref{fig:cp_small} and \ref{fig:cB_small}. The supersonic case is shown in Figure \ref{fig:saxplot_supersonic}. There is appreciable time variation, but all timesteps follow the patterns seen in the mean. The mean monopole $\ell = 0$ starts negative and then tends to zero, while the mean dipole $\ell = 1$ starts positive, falls to be negative, and then tends to zero. The higher multipoles $\ell \geq 2$ start positive and then tend to zero. The lower multipoles $\ell \leq 1$ show the most time-variation at low bin number, around their most negative values, but the higher multipoles show the most variation at high bin number, when they tend to zero. Figure \ref{fig:saxplot_transsonic} shows the transonic case. Here the lower multipoles start negative and then tend to zero, again showing the most variation at low bin number. The higher multipoles fall until bin 5 and then rise towards zero. The variation is quite significant in the highest multipoles $\ell \geq 4$, making them lower in signal-to-noise. Figure \ref{fig:saxplot_subsonic} shows a sub-Alfv\'enic and subsonic case, which behaves similarly in the lower multipoles. $\ell=2,3$ fall until bin 3 and then rise to be consistent with zero while $\ell=4,5$ fall until bin 5, cross zero at bin 10 and then fall to zero.

\added{
\section{Convergence Tests}
\label{sec:convergence}
We now consider whether our statistic is stable to a change in the resolution of the simulations. We take the most ``extreme'' conditions we study, i.e. those most likely to form highly localized, challenging-to-resolve density structures such as shocks, and study it at a resolution of $256^3$ and $512^3$. We perform this test on a simulation with $M_S =7$, $M_A =1.4$. Because the 3PCF shows some time variation, we use snapshots from multiple timesteps: 9 snapshots from the $256^3$ simulation, and 3 from the $512^3$ simulation. We use fewer of the $512^3$ snapshots, due to their 3PCFs' greater computational expense. We compare the 3PCF diagonal and half-diagonal for $256^3$ and $512^3$ in Figures \ref{fig:resolution_diag} and \ref{fig:resolution_halfdiag}. While a precise comparison is difficult due to the time-variability of the 3PCF, the time-averaged 3PCF appears to be well converged for the lower multipoles $\ell=0,1$, with both resolutions agreeing within one standard deviation. At the higher multipoles, while the two resolutions disagree somewhat at low bin number, they still follow similar trends. Convergence is expected to be harder for higher multipoles, as smaller angular scales are probed and these may be more strongly affected by numerical noise.

\begin{figure}
\includegraphics[width=1.10\columnwidth]{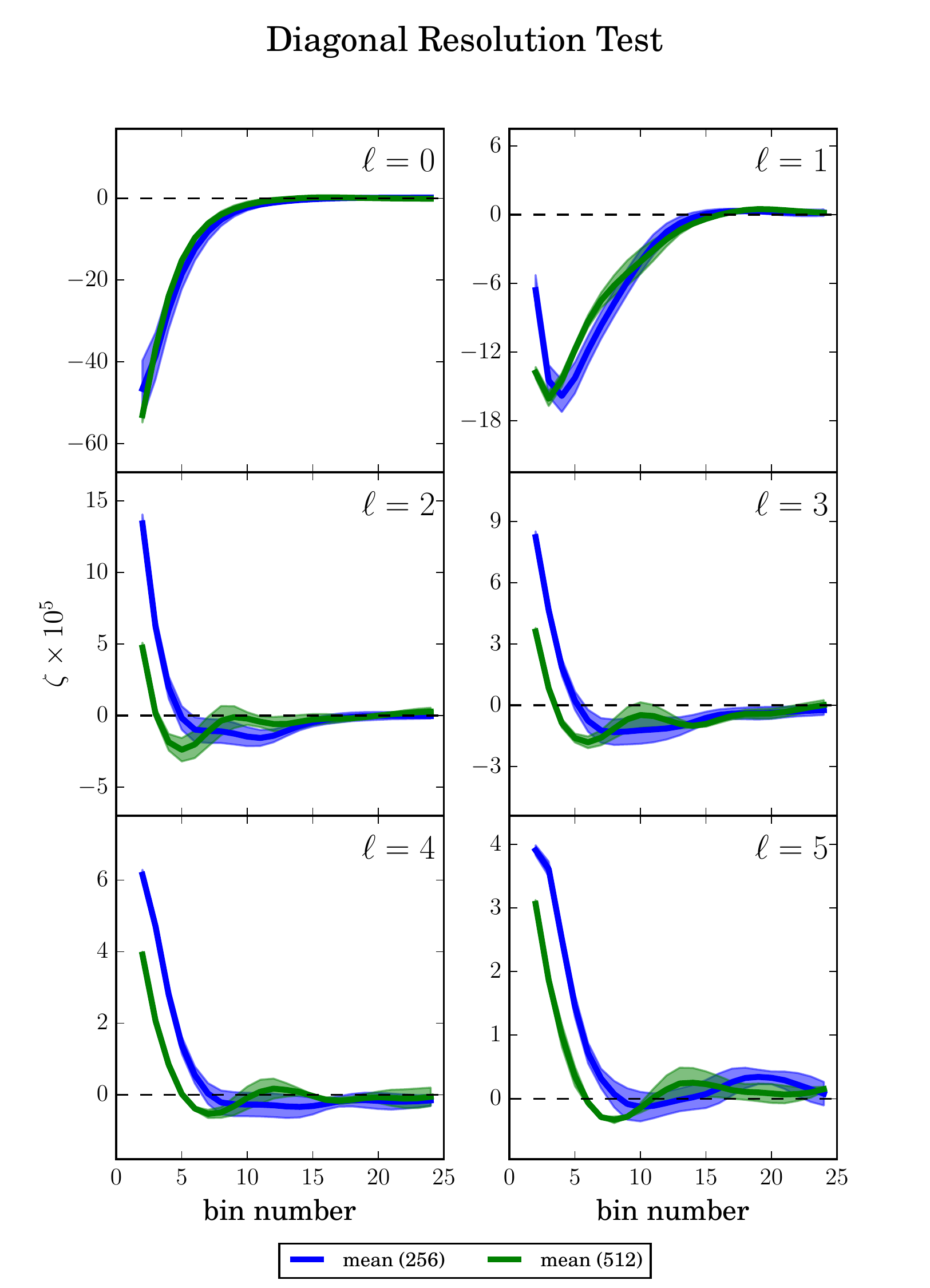}
\caption{These panels show the diagonals of the 3PCF multipoles from bin 0 to 25 averaged over nine time steps for the $256^3$ simulation (blue line) and averaged over three time steps for the $512^3$ simulation (green). The colored bands show the standard deviation of the mean, which accounts for the differing number of time steps used to determine each mean. The 3PCF shows similar trends in both resolutions.}
\label{fig:resolution_diag}
\end{figure}

\begin{figure}
\includegraphics[width=1.10\columnwidth]{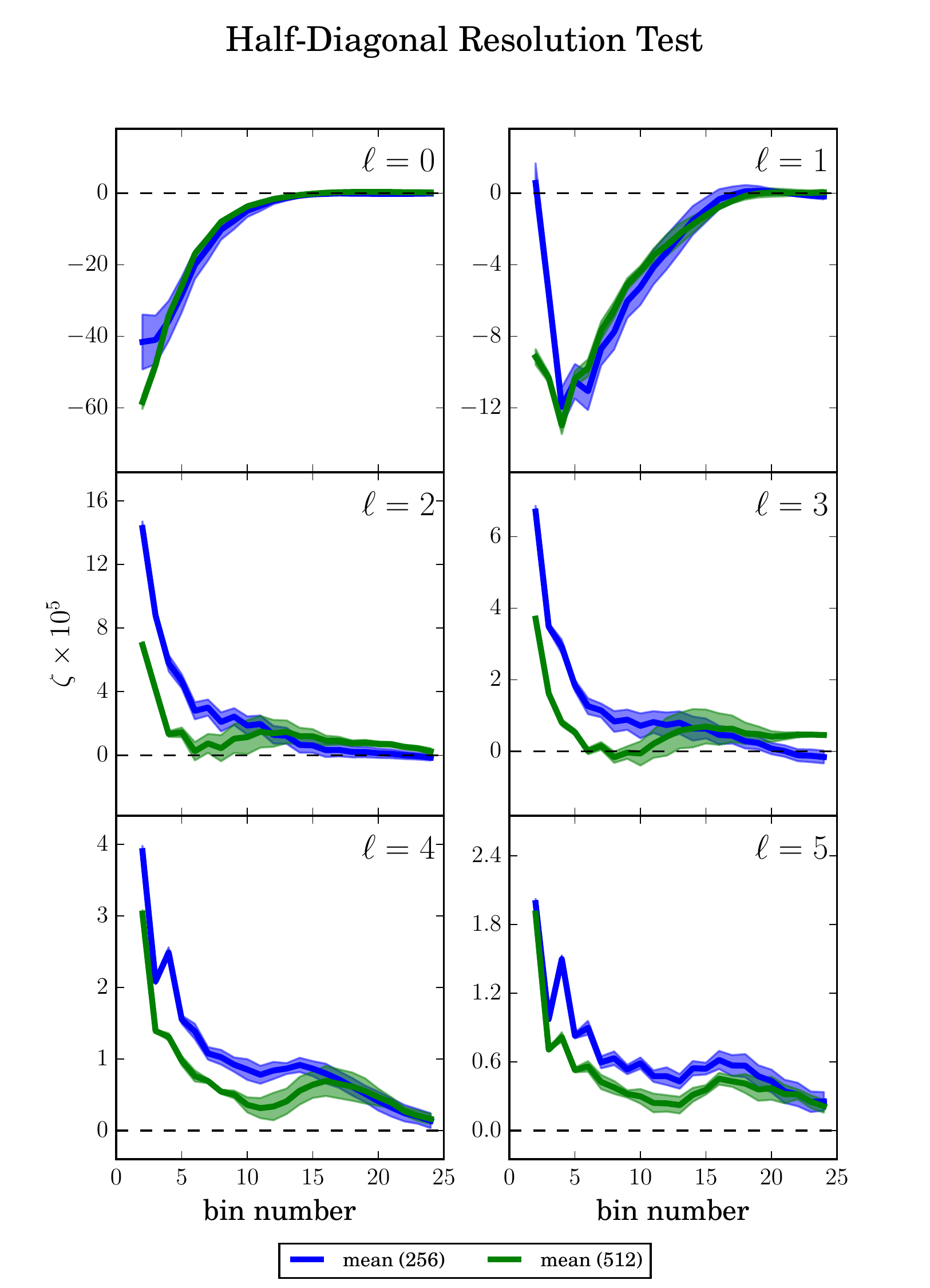}
\caption{Similar to Figure \ref{fig:resolution_diag}, but for the 3PCF half-diagonal. Again, the 3PCF shows similar trends in both resolutions.}
\label{fig:resolution_halfdiag}
\end{figure}

}

\section{Discussion \& Conclusions}
\label{sec:concs}

Quantifying the spatial structure of the ISM with statistical diagnostics allows simulations to be more directly compared with observations. The power spectrum probes the turbulent power cascade as a function of scale, but does not capture phase information which is important in describing ISM structure. The bispectrum captures phase information but because of its computational cost (scaling as $N^3$, with $N$ the number of wave-vectors), previous studies have only studied the angle-averaged bispectrum. We implement a fast algorithm for the angle-dependent 3PCF 
(scaling as $N_g \log N_g$, with $N_g$ the number of grid cells), the configuration-space analog of the bispectrum. The speed of this algorithm makes the angle-dependent 3PCF feasible to calculate on simulations and observations, enabling the 3PCF to be used as a statistical diagnostic that probes phase information.

Previous work \citep{Burkhart09} has shown that the angle-averaged-bispectrum of the density field is sensitive to the sonic and Alfv\'enic Mach numbers, which are important plasma parameters that are not always straightforward to measure in astrophysical observations. The 3PCF is a natural statistic to use for turbulence studies, and our fast implementation makes it a practical statistic to use.

To demonstrate our 3PCF algorithm, we run it on a suite of MHD turbulence simulations with a wide range of sonic and Alfv\'enic Mach numbers. We find that the 3PCF is time-stable and that it is sensitive to both sonic and Alfv\'enic Mach number. The higher multipoles of the 3PCF, capturing its angular dependence, show stronger sensitivity to the simulation conditions than does the angle-averaged 3PCF (i.e. the 3PCF monopole), demonstrating that the angular dependence of the 3PCF provides useful information which was not captured in the earlier works on the angle-averaged-bispectrum of turbulent density fields. Much of the 3PCF's discriminatory power is retained using simple compressions of its side-length dependence such as the diagonal ($r_2=r_1$) and half-diagonal ($r_2 =  r_1/2$); see also \cite{burkhart15} for work on similar compressions of the bispectrum. These correspond to isosceles and $2:1$ triangles. We leave a study of the ideal compression of the 3PCF for future work.

This work shows that the 3PCF is a useful statistical diagnostic for MHD turbulence simulations, but we have not investigated the impact of observational effects. In particular, observations are often integrated over the line-of-sight, yielding a 2-D column density map rather than a 3-D density cube. We leave a study of the 3PCF of column density maps for future work; however we note that the angle-averaged bispectrum of column density maps has been investigated in \cite{Burkhart09}. That work found that the 3-D density and 2-D column density bispectrum were similar, though the 2-D bispectrum suffers from smearing out of coherent structures such as filaments due to the line-of-sight integration. We therefore might expect a similar behavior with the 3PCF.

Another promising direction for future work is calculating the anisotropic 3PCF of MHD turbulence simulations. The isotropic, or rotation-averaged, 3PCF measured in this work depends only on the side lengths and opening angle of triangles of points, and not the overall orientation of the triangle. However, the magnetic field imposes a preferred direction. \cite{Slepian_aniso_alg17} present an algorithm for calculating the anisotropic 3PCF of galaxies, where the 3PCF is assumed to depend on the angles of the triangle sides with respect to some preferred direction (in the case of redshift space distortions in cosmology, the line-of-sight is a preferred direction). This algorithm also scales as $N^2$, accelerating to $N_{\rm g} \log N_{\rm g}$ with FFTs, and \cite{Galactos17} present a parallel, high-performance, scalable implementation of the $N^2$ version. The anisotropic 3PCF of MHD turbulence using the magnetic field direction as the preferred direction could show even greater sensitivity to Alfv\'enic Mach number than the isotropic 3PCF used in this work.

\acknowledgments
SKNP acknowledges support from a Sir James Lougheed Award of Distinction and a Natural Sciences and Engineering Research Council of Canada Postgraduate Scholarship (held previously to the Lougheed). Support for this work was also provided by the National Aeronautics and Space Administration through Einstein Postdoctoral Fellowship Award Number PF7-180167 issued by the Chandra X-ray Observatory Center, which is operated by the Smithsonian Astrophysical Observatory for and on behalf of the National Aeronautics Space Administration under contract NAS8-03060.  ZS also acknowledges support from a Chamberlain Fellowship at Lawrence Berkeley National Laboratory (held previously to the Einstein) and from the Berkeley Center for Cosmological Physics. ZS thanks Alex Krolewski and Alexandre Lazarian for many useful conversations. BB acknowledges the generous support of the Institute for Theory and Computation (ITC) Postdoctoral Fellowship program and the Sub-Millimeter Array (SMA) Postdoctoral Fellowship program. The computations in this paper were run on the Odyssey cluster supported by the FAS Division of Science, Research Computing Group at Harvard University.


\bibliographystyle{yahapj}
\bibliography{refs.bib,morerefs.bib}

\begin{thebibliography}{}
\providecommand\natexlab[1]{#1}
\providecommand\JournalTitle[1]{#1}

\bibitem[{{Armstrong} {et~al.}(1995){Armstrong}, {Rickett}, \&
  {Spangler}}]{Armstrong95}
{Armstrong}, J.~W., {Rickett}, B.~J., \& {Spangler}, S.~R. 1995,
  \href{http://dx.doi.org/10.1086/175515}{\JournalTitle{\apj}, 443, 209}

\bibitem[{{Boldyrev}(2006)}]{Boldyrev}
{Boldyrev}, S. 2006,
  \href{http://dx.doi.org/10.1103/PhysRevLett.96.115002}{\JournalTitle{Physical
  Review Letters}, 96, 115002}

\bibitem[{{Burkhart} {et~al.}(2009){Burkhart}, {Falceta-Gon{\c c}alves},
  {Kowal}, \& {Lazarian}}]{Burkhart09}
{Burkhart}, B., {Falceta-Gon{\c c}alves}, D., {Kowal}, G., \& {Lazarian}, A.
  2009,
  \href{http://dx.doi.org/10.1088/0004-637X/693/1/250}{\JournalTitle{\apj},
  693, 250}

\bibitem[{{Burkhart} \& {Lazarian}(2012)}]{Burkhart12}
{Burkhart}, B., \& {Lazarian}, A. 2012,
  \href{http://dx.doi.org/10.1088/2041-8205/755/1/L19}{\JournalTitle{\apjl},
  755, L19}

\bibitem[{Burkhart \& Lazarian(2012)}]{Burkhart2012}
Burkhart, B., \& Lazarian, A. 2012,
  \href{http://dx.doi.org/10.1088/2041-8205/755/1/L19}{\JournalTitle{ApJ}, 755,
  L19}

\bibitem[{{Burkhart} \& {Lazarian}(2016)}]{Burkhart2016ApJ...827...26B}
{Burkhart}, B., \& {Lazarian}, A. 2016,
  \href{http://dx.doi.org/10.3847/0004-637X/827/1/26}{\JournalTitle{\apj}, 827,
  26}

\bibitem[{{Burkhart} {et~al.}(2013{\natexlab{a}}){Burkhart}, {Lazarian},
  {Goodman}, \& {Rosolowsky}}]{burkhart13a}
{Burkhart}, B., {Lazarian}, A., {Goodman}, A., \& {Rosolowsky}, E.
  2013{\natexlab{a}},
  \href{http://dx.doi.org/10.1088/0004-637X/770/2/141}{\JournalTitle{\apj},
  770, 141}

\bibitem[{{Burkhart} {et~al.}(2014){Burkhart}, {Lazarian}, {Le{\~a}o}, {de
  Medeiros}, \& {Esquivel}}]{burkhart14}
{Burkhart}, B., {Lazarian}, A., {Le{\~a}o}, I.~C., {de Medeiros}, J.~R., \&
  {Esquivel}, A. 2014,
  \href{http://dx.doi.org/10.1088/0004-637X/790/2/130}{\JournalTitle{\apj},
  790, 130}

\bibitem[{{Burkhart} {et~al.}(2013{\natexlab{b}}){Burkhart}, {Lazarian},
  {Ossenkopf}, \& {Stutzki}}]{Bur13}
{Burkhart}, B., {Lazarian}, A., {Ossenkopf}, V., \& {Stutzki}, J.
  2013{\natexlab{b}},
  \href{http://dx.doi.org/10.1088/0004-637X/771/2/123}{\JournalTitle{\apj},
  771, 123}

\bibitem[{{Burkhart} {et~al.}(2015){Burkhart}, {Lee}, {Murray}, \&
  {Stanimirovic}}]{burkhart15}
{Burkhart}, B., {Lee}, M.~Y., {Murray}, C., \& {Stanimirovic}, S. 2015,
  \JournalTitle{\apj}, 811, 28

\bibitem[{{Burkhart} {et~al.}(2010){Burkhart}, {Stanimirovi{\'c}}, {Lazarian},
  \& {Kowal}}]{burkhart10}
{Burkhart}, B., {Stanimirovi{\'c}}, S., {Lazarian}, A., \& {Kowal}, G. 2010,
  \href{http://dx.doi.org/10.1088/0004-637X/708/2/1204}{\JournalTitle{\apj},
  708, 1204}

\bibitem[{{Cardelli} {et~al.}(1989){Cardelli}, {Clayton}, \&
  {Mathis}}]{Cardelli89}
{Cardelli}, J.~A., {Clayton}, G.~C., \& {Mathis}, J.~S. 1989,
  \href{http://dx.doi.org/10.1086/167900}{\JournalTitle{\apj}, 345, 245}

\bibitem[{{Chen} {et~al.}(2017){Chen}, {Burkhart}, {Goodman}, \&
  {Collins}}]{chen2017}
{Chen}, H., {Burkhart}, B., {Goodman}, A.~A., \& {Collins}, D.~C. 2017,
  \JournalTitle{ArXiv e-prints},
  \href{http://arxiv.org/abs/1707.09356}{{\sffamily arXiv:1707.09356}}

\bibitem[{{Chepurnov} {et~al.}(2015){Chepurnov}, {Burkhart}, {Lazarian}, \&
  {Stanimirovic}}]{chepurnov15}
{Chepurnov}, A., {Burkhart}, B., {Lazarian}, A., \& {Stanimirovic}, S. 2015,
  \href{http://dx.doi.org/10.1088/0004-637X/810/1/33}{\JournalTitle{\apj}, 810,
  33}

\bibitem[{{Chepurnov} {et~al.}(2008){Chepurnov}, {Gordon}, {Lazarian}, \&
  {Stanimirovic}}]{Chepurnov2008}
{Chepurnov}, A., {Gordon}, J., {Lazarian}, A., \& {Stanimirovic}, S. 2008,
  \href{http://dx.doi.org/10.1086/591655}{\JournalTitle{\apj}, 688, 1021}

\bibitem[{{Chepurnov} \& {Lazarian}(2010)}]{CheL10}
{Chepurnov}, A., \& {Lazarian}, A. 2010,
  \href{http://dx.doi.org/10.1088/0004-637X/710/1/853}{\JournalTitle{\apj},
  710, 853}

\bibitem[{Cho \& Lazarian(2002)}]{Cho2002}
Cho, J., \& Lazarian, A. 2002,
  \href{http://dx.doi.org/10.1103/PhysRevLett.88.245001}{\JournalTitle{PhRvL},
  88}, \href{http://arxiv.org/abs/0205282}{{\sffamily arXiv:0205282
  [astro-ph]}}

\bibitem[{Cho \& Lazarian(2003)}]{Cho2003}
---. 2003,
  \href{http://dx.doi.org/10.1046/j.1365-8711.2003.06941.x}{\JournalTitle{MNRAS},
  345, 325}

\bibitem[{{Cho} \& {Lazarian}(2003{\natexlab{a}})}]{cho03}
{Cho}, J., \& {Lazarian}, A. 2003{\natexlab{a}},
  \href{http://dx.doi.org/10.1046/j.1365-8711.2003.06941.x}{\JournalTitle{\mnras},
  345, 325}

\bibitem[{{Cho} \& {Lazarian}(2003{\natexlab{b}})}]{CL03}
---. 2003{\natexlab{b}},
  \href{http://dx.doi.org/10.1046/j.1365-8711.2003.06941.x}{\JournalTitle{\mnras},
  345, 325}

\bibitem[{{Cho} \& {Lazarian}(2009)}]{Cho2009}
---. 2009,
  \href{http://dx.doi.org/10.1088/0004-637X/701/1/236}{\JournalTitle{\apj},
  701, 236}

\bibitem[{{Collins} {et~al.}(2012){Collins}, {Kritsuk}, {Padoan}, {Li}, {Xu},
  {Ustyugov}, \& {Norman}}]{Collins12a}
{Collins}, D.~C., {Kritsuk}, A.~G., {Padoan}, P., {et~al.} 2012,
  \href{http://dx.doi.org/10.1088/0004-637X/750/1/13}{\JournalTitle{\apj}, 750,
  13}

\bibitem[{{Corasaniti}(2006)}]{Corasaniti06}
{Corasaniti}, P.~S. 2006,
  \href{http://dx.doi.org/10.1111/j.1365-2966.2006.10825.x}{\JournalTitle{\mnras},
  372, 191}

\bibitem[{{Dalal} {et~al.}(2008){Dalal}, {Dor{\'e}}, {Huterer}, \&
  {Shirokov}}]{Dalal08}
{Dalal}, N., {Dor{\'e}}, O., {Huterer}, D., \& {Shirokov}, A. 2008,
  \href{http://dx.doi.org/10.1103/PhysRevD.77.123514}{\JournalTitle{\prd}, 77,
  123514}

\bibitem[{{Desjacques} {et~al.}(2016){Desjacques}, {Jeong}, \&
  {Schmidt}}]{Desjacques16}
{Desjacques}, V., {Jeong}, D., \& {Schmidt}, F. 2016, \JournalTitle{ArXiv
  e-prints}, \href{http://arxiv.org/abs/1611.09787}{{\sffamily
  arXiv:1611.09787}}

\bibitem[{{Elmegreen} \& {Scalo}(2004)}]{ElmegreenScalo}
{Elmegreen}, B.~G., \& {Scalo}, J. 2004,
  \href{http://dx.doi.org/10.1146/annurev.astro.41.011802.094859}{\JournalTitle{\araa},
  42, 211}

\bibitem[{{Esquivel} \& {Lazarian}(2005)}]{esquivel05}
{Esquivel}, A., \& {Lazarian}, A. 2005,
  \href{http://dx.doi.org/10.1086/432458}{\JournalTitle{\apj}, 631, 320}

\bibitem[{{Esquivel} \& {Lazarian}(2011)}]{Esquivel11}
---. 2011,
  \href{http://dx.doi.org/10.1088/0004-637X/740/2/117}{\JournalTitle{\apj},
  740, 117}

\bibitem[{Federrath(2013)}]{doi:10.1093/mnras/stt1644}
Federrath, C. 2013,
  \href{http://dx.doi.org/10.1093/mnras/stt1644}{\JournalTitle{Monthly Notices
  of the Royal Astronomical Society}, 436, 1245}

\bibitem[{Federrath(2016)}]{federrath_2016}
---. 2016,
  \href{http://dx.doi.org/10.1017/S0022377816001069}{\JournalTitle{Journal of
  Plasma Physics}, 82}

\bibitem[{{Federrath}(2016)}]{2016MNRAS.457..375F}
{Federrath}, C. 2016,
  \href{http://dx.doi.org/10.1093/mnras/stv2880}{\JournalTitle{\mnras}, 457,
  375}

\bibitem[{Federrath {et~al.}(2008)Federrath, Klessen, \&
  Schmidt}]{Federrath2008}
Federrath, C., Klessen, R.~S., \& Schmidt, W. 2008,
  \href{http://dx.doi.org/10.1086/595280}{\JournalTitle{ApJ}, 688, L79}

\bibitem[{{Federrath} {et~al.}(2009){Federrath}, {Klessen}, \&
  {Schmidt}}]{Federrath09a}
{Federrath}, C., {Klessen}, R.~S., \& {Schmidt}, W. 2009,
  \href{http://dx.doi.org/10.1088/0004-637X/692/1/364}{\JournalTitle{\apj},
  692, 364}

\bibitem[{{Federrath} {et~al.}(2010){Federrath}, {Roman-Duval}, {Klessen},
  {Schmidt}, \& {Mac Low}}]{2010A&A...512A..81F}
{Federrath}, C., {Roman-Duval}, J., {Klessen}, R.~S., {Schmidt}, W., \& {Mac
  Low}, M.-M. 2010,
  \href{http://dx.doi.org/10.1051/0004-6361/200912437}{\JournalTitle{\aap},
  512, A81}

\bibitem[{{Friesen} {et~al.}(2017){Friesen}, {Patwary}, {Austin}, {Satish},
  {Slepian}, {Sundaram}, {Bard}, {Eisenstein}, {Deslippe}, {Dubey}, \&
  {Prabhat}}]{Galactos17}
{Friesen}, B., {Patwary}, M.~M.~A., {Austin}, B., {et~al.} 2017,
  \JournalTitle{ArXiv e-prints},
  \href{http://arxiv.org/abs/1709.00086}{{\sffamily arXiv:1709.00086}}

\bibitem[{{Gaensler} {et~al.}(2011){Gaensler}, {Haverkorn}, {Burkhart},
  {Newton-McGee}, {Ekers}, {Lazarian}, {McClure-Griffiths}, {Robishaw},
  {Dickey}, \& {Green}}]{Gaensler2011}
{Gaensler}, B.~M., {Haverkorn}, M., {Burkhart}, B., {et~al.} 2011,
  \href{http://dx.doi.org/10.1038/nature10446}{\JournalTitle{\nat}, 478, 214}

\bibitem[{{Goldreich} \& {Sridhar}(1995)}]{Goldreich95a}
{Goldreich}, P., \& {Sridhar}, S. 1995,
  \href{http://dx.doi.org/10.1086/175121}{\JournalTitle{\apj}, 438, 763}

\bibitem[{{Goodman} {et~al.}(2009){Goodman}, {Rosolowsky}, {Borkin}, {Foster},
  {Halle}, {Kauffmann}, \& {Pineda}}]{goodman09}
{Goodman}, A.~A., {Rosolowsky}, E.~W., {Borkin}, M.~A., {et~al.} 2009,
  \href{http://dx.doi.org/10.1038/nature07609}{\JournalTitle{\nat}, 457, 63}

\bibitem[{{Heyer} {et~al.}(2009){Heyer}, {Krawczyk}, {Duval}, \&
  {Jackson}}]{Heyer09a}
{Heyer}, M., {Krawczyk}, C., {Duval}, J., \& {Jackson}, J.~M. 2009,
  \href{http://dx.doi.org/10.1088/0004-637X/699/2/1092}{\JournalTitle{\apj},
  699, 1092}

\bibitem[{{Heyer} \& {Brunt}(2004)}]{Heyer04a}
{Heyer}, M.~H., \& {Brunt}, C.~M. 2004,
  \href{http://dx.doi.org/10.1086/425978}{\JournalTitle{\apjl}, 615, L45}

\bibitem[{{Heyer} \& {Brunt}(2012)}]{Hey12}
---. 2012,
  \href{http://dx.doi.org/10.1111/j.1365-2966.2011.20142.x}{\JournalTitle{\mnras},
  420, 1562}

\bibitem[{{Hill} {et~al.}(2008){Hill}, {Benjamin}, {Kowal}, {Reynolds},
  {Haffner}, \& {Lazarian}}]{Hill2008}
{Hill}, A.~S., {Benjamin}, R.~A., {Kowal}, G., {et~al.} 2008,
  \href{http://dx.doi.org/10.1086/590543}{\JournalTitle{\apj}, 686, 363}

\bibitem[{{Kainulainen} {et~al.}(2009){Kainulainen}, {Beuther}, {Henning}, \&
  {Plume}}]{Kainulainen09a}
{Kainulainen}, J., {Beuther}, H., {Henning}, T., \& {Plume}, R. 2009,
  \href{http://dx.doi.org/10.1051/0004-6361/200913605}{\JournalTitle{\aap},
  508, L35}

\bibitem[{{Konstandin} {et~al.}(2012){Konstandin}, {Federrath}, {Klessen}, \&
  {Schmidt}}]{2012JFM...692..183K}
{Konstandin}, L., {Federrath}, C., {Klessen}, R.~S., \& {Schmidt}, W. 2012,
  \href{http://dx.doi.org/10.1017/jfm.2011.503}{\JournalTitle{Journal of Fluid
  Mechanics}, 692, 183}

\bibitem[{{Kowal} \& {Lazarian}(2007)}]{Kowal07}
{Kowal}, G., \& {Lazarian}, A. 2007,
  \href{http://dx.doi.org/10.1086/521788}{\JournalTitle{\apjl}, 666, L69}

\bibitem[{{Kowal} \& {Lazarian}(2010{\natexlab{a}})}]{Kowal10}
---. 2010{\natexlab{a}},
  \href{http://dx.doi.org/10.1088/0004-637X/720/1/742}{\JournalTitle{\apj},
  720, 742}

\bibitem[{{Kowal} \& {Lazarian}(2010{\natexlab{b}})}]{KowL10}
---. 2010{\natexlab{b}},
  \href{http://dx.doi.org/10.1088/0004-637X/720/1/742}{\JournalTitle{\apj},
  720, 742}

\bibitem[{{Kowal} {et~al.}(2007){Kowal}, {Lazarian}, \& {Beresnyak}}]{Kowal07a}
{Kowal}, G., {Lazarian}, A., \& {Beresnyak}, A. 2007,
  \href{http://dx.doi.org/10.1086/511515}{\JournalTitle{\apj}, 658, 423}

\bibitem[{Kowal {et~al.}(2007)Kowal, Lazarian, \& Beresnyak}]{Kowal2007}
Kowal, G., Lazarian, A., \& Beresnyak, A. 2007,
  \href{http://dx.doi.org/10.1086/511515}{\JournalTitle{ApJ}, 658, 423}

\bibitem[{Kowal {et~al.}(2009)Kowal, Lazarian, Vishniac, \&
  Otmianowska-Mazur}]{Kowal2009}
Kowal, G., Lazarian, A., Vishniac, E.~T., \& Otmianowska-Mazur, K. 2009,
  \href{http://dx.doi.org/10.1088/0004-637X/700/1/63}{\JournalTitle{ApJ}, 700,
  63}

\bibitem[{Kowal {et~al.}(2011)Kowal, Pino, \& Lazarian}]{Kowal2011}
Kowal, G., Pino, E. M. d. G.~D., \& Lazarian, A. 2011,
  \href{http://dx.doi.org/10.1088/0004-637X/735/2/102}{\JournalTitle{ApJ},
  735}, \href{http://arxiv.org/abs/1103.2984}{{\sffamily arXiv:1103.2984}}

\bibitem[{Kritsuk {et~al.}(2007)Kritsuk, Norman, Padoan, \&
  Wagner}]{0004-637X-665-1-416}
Kritsuk, A.~G., Norman, M.~L., Padoan, P., \& Wagner, R. 2007,
  \href{http://stacks.iop.org/0004-637X/665/i=1/a=416}{\JournalTitle{The
  Astrophysical Journal}, 665, 416}

\bibitem[{{Larson}(1981)}]{Larson81a}
{Larson}, R.~B. 1981, \JournalTitle{\mnras}, 194, 809

\bibitem[{{Lazarian} \& {Beresnyak}(2006)}]{lazarianberesnyak06}
{Lazarian}, A., \& {Beresnyak}, A. 2006,
  \href{http://dx.doi.org/10.1111/j.1365-2966.2006.11093.x}{\JournalTitle{\mnras},
  373, 1195}

\bibitem[{{Lazarian} \& {Pogosyan}(2004)}]{LP04}
{Lazarian}, A., \& {Pogosyan}, D. 2004,
  \href{http://dx.doi.org/10.1086/422462}{\JournalTitle{\apj}, 616, 943}

\bibitem[{{Lazarian} \& {Pogosyan}(2006)}]{LP06}
---. 2006, \href{http://dx.doi.org/10.1086/508012}{\JournalTitle{\apj}, 652,
  1348}

\bibitem[{{Lazarian} \& {Pogosyan}(2008)}]{LP08}
---. 2008, \href{http://dx.doi.org/10.1086/591238}{\JournalTitle{\apj}, 686,
  350}

\bibitem[{{McKee} \& {Ostriker}(2007)}]{Mckee_Ostriker2007}
{McKee}, C.~F., \& {Ostriker}, E.~C. 2007,
  \href{http://dx.doi.org/10.1146/annurev.astro.45.051806.110602}{\JournalTitle{\araa},
  45, 565}

\bibitem[{{Omukai}(2000)}]{Omukai00a}
{Omukai}, K. 2000, \href{http://dx.doi.org/10.1086/308776}{\JournalTitle{\apj},
  534, 809}

\bibitem[{Pippig(2013)}]{10.1137/120885887}
Pippig, M. 2013, \JournalTitle{SIAM Journal on Scientific Computing}, 35, C213

\bibitem[{{Planck Collaboration} {et~al.}(2016{\natexlab{a}}){Planck
  Collaboration}, {Aghanim}, {Alves}, {Arzoumanian}, {Aumont}, {Baccigalupi},
  {Ballardini}, {Banday}, {Barreiro}, {Bartolo}, {Basak}, {Benabed}, {Bernard},
  {Bersanelli}, {Bielewicz}, {Bonavera}, {Bond}, {Borrill}, {Bouchet},
  {Boulanger}, {Bracco}, {Bucher}, {Burigana}, {Calabrese}, {Cardoso},
  {Chiang}, {Colombo}, {Combet}, {Comis}, {Couchot}, {Coulais}, {Crill},
  {Curto}, {Cuttaia}, {Davis}, {de Bernardis}, {de Rosa}, {de Zotti},
  {Delabrouille}, {Delouis}, {Di Valentino}, {Dickinson}, {Diego}, {Dor{\'e}},
  {Douspis}, {Ducout}, {Dupac}, {Dusini}, {Efstathiou}, {Elsner}, {En{\ss}lin},
  {Eriksen}, {Falgarone}, {Fantaye}, {Ferri{\`e}re}, {Finelli}, {Frailis},
  {Fraisse}, {Franceschi}, {Frolov}, {Galeotta}, {Galli}, {Ganga},
  {G{\'e}nova-Santos}, {Gerbino}, {Ghosh}, {Gonz{\'a}lez-Nuevo}, {G{\'o}rski},
  {Gratton}, {Gregorio}, {Gruppuso}, {Gudmundsson}, {Guillet}, {Hansen},
  {Helou}, {Henrot-Versill{\'e}}, {Herranz}, {Hivon}, {Huang}, {Jaffe},
  {Jaffe}, {Jones}, {Keih{\"a}nen}, {Keskitalo}, {Kisner}, {Krachmalnicoff},
  {Kunz}, {Kurki-Suonio}, {Lagache}, {L{\"a}hteenm{\"a}ki}, {Lamarre},
  {Langer}, {Lasenby}, {Lattanzi}, {Le Jeune}, {Levrier}, {Liguori}, {Lilje},
  {L{\'o}pez-Caniego}, {Lubin}, {Mac{\'{\i}}as-P{\'e}rez}, {Maggio}, {Maino},
  {Mandolesi}, {Mangilli}, {Maris}, {Martin}, {Mart{\'{\i}}nez-Gonz{\'a}lez},
  {Matarrese}, {Mauri}, {McEwen}, {Melchiorri}, {Mennella}, {Migliaccio},
  {Miville-Desch{\^e}nes}, {Molinari}, {Moneti}, {Montier}, {Morgante}, {Moss},
  {Naselsky}, {Natoli}, {Neveu}, {N{\o}rgaard-Nielsen}, {Oppermann},
  {Oxborrow}, {Pagano}, {Paoletti}, {Partridge}, {Perdereau}, {Perotto},
  {Pettorino}, {Piacentini}, {Plaszczynski}, {Polenta}, {Rachen}, {Rebolo},
  {Reinecke}, {Remazeilles}, {Renzi}, {Ristorcelli}, {Rocha}, {Rossetti},
  {Roudier}, {Ruiz-Granados}, {Salvati}, {Sandri}, {Savelainen}, {Scott},
  {Sirignano}, {Soler}, {Suur-Uski}, {Tauber}, {Tavagnacco}, {Tenti},
  {Toffolatti}, {Tomasi}, {Tristram}, {Trombetti}, {Valiviita}, {Vansyngel},
  {Van Tent}, {Vielva}, {Villa}, {Wandelt}, {Wehus}, {Zacchei}, \&
  {Zonca}}]{PlanckXLIV}
{Planck Collaboration}, {Aghanim}, N., {Alves}, M.~I.~R., {et~al.}
  2016{\natexlab{a}},
  \href{http://dx.doi.org/10.1051/0004-6361/201628636}{\JournalTitle{\aap},
  596, A105}

\bibitem[{{Planck Collaboration} {et~al.}(2016{\natexlab{b}}){Planck
  Collaboration}, {Ade}, {Aghanim}, {Alves}, {Aniano}, {Arnaud}, {Ashdown},
  {Aumont}, {Baccigalupi}, {Banday}, {Barreiro}, {Bartolo}, {Battaner},
  {Benabed}, {Benoit-L{\'e}vy}, {Bernard}, {Bersanelli}, {Bielewicz},
  {Bonaldi}, {Bonavera}, {Bond}, {Borrill}, {Bouchet}, {Boulanger}, {Burigana},
  {Butler}, {Calabrese}, {Cardoso}, {Catalano}, {Chamballu}, {Chiang},
  {Christensen}, {Clements}, {Colombi}, {Colombo}, {Couchot}, {Crill}, {Curto},
  {Cuttaia}, {Danese}, {Davies}, {Davis}, {de Bernardis}, {de Rosa}, {de
  Zotti}, {Delabrouille}, {Dickinson}, {Diego}, {Dole}, {Donzelli}, {Dor{\'e}},
  {Douspis}, {Draine}, {Ducout}, {Dupac}, {Efstathiou}, {Elsner}, {En{\ss}lin},
  {Eriksen}, {Falgarone}, {Finelli}, {Forni}, {Frailis}, {Fraisse},
  {Franceschi}, {Frejsel}, {Galeotta}, {Galli}, {Ganga}, {Ghosh}, {Giard},
  {Gjerl{\o}w}, {Gonz{\'a}lez-Nuevo}, {G{\'o}rski}, {Gregorio}, {Gruppuso},
  {Guillet}, {Hansen}, {Hanson}, {Harrison}, {Henrot-Versill{\'e}},
  {Hern{\'a}ndez-Monteagudo}, {Herranz}, {Hildebrandt}, {Hivon}, {Holmes},
  {Hovest}, {Huffenberger}, {Hurier}, {Jaffe}, {Jaffe}, {Jones},
  {Keih{\"a}nen}, {Keskitalo}, {Kisner}, {Kneissl}, {Knoche}, {Kunz},
  {Kurki-Suonio}, {Lagache}, {Lamarre}, {Lasenby}, {Lattanzi}, {Lawrence},
  {Leonardi}, {Levrier}, {Liguori}, {Lilje}, {Linden-V{\o}rnle},
  {L{\'o}pez-Caniego}, {Lubin}, {Mac{\'{\i}}as-P{\'e}rez}, {Maffei}, {Maino},
  {Mandolesi}, {Maris}, {Marshall}, {Martin}, {Mart{\'{\i}}nez-Gonz{\'a}lez},
  {Masi}, {Matarrese}, {Mazzotta}, {Melchiorri}, {Mendes}, {Mennella},
  {Migliaccio}, {Miville-Desch{\^e}nes}, {Moneti}, {Montier}, {Morgante},
  {Mortlock}, {Munshi}, {Murphy}, {Naselsky}, {Natoli}, {N{\o}rgaard-Nielsen},
  {Novikov}, {Novikov}, {Oxborrow}, {Pagano}, {Pajot}, {Paladini}, {Paoletti},
  {Pasian}, {Perdereau}, {Perotto}, {Perrotta}, {Pettorino}, {Piacentini},
  {Piat}, {Plaszczynski}, {Pointecouteau}, {Polenta}, {Ponthieu}, {Popa},
  {Pratt}, {Prunet}, {Puget}, {Rachen}, {Reach}, {Rebolo}, {Reinecke},
  {Remazeilles}, {Renault}, {Ristorcelli}, {Rocha}, {Roudier},
  {Rubi{\~n}o-Mart{\'{\i}}n}, {Rusholme}, {Sandri}, {Santos}, {Scott},
  {Spencer}, {Stolyarov}, {Sudiwala}, {Sunyaev}, {Sutton}, {Suur-Uski},
  {Sygnet}, {Tauber}, {Terenzi}, {Toffolatti}, {Tomasi}, {Tristram}, {Tucci},
  {Umana}, {Valenziano}, {Valiviita}, {Van Tent}, {Vielva}, {Villa}, {Wade},
  {Wandelt}, {Wehus}, {Ysard}, {Yvon}, {Zacchei}, \& {Zonca}}]{PlanckXXIX}
{Planck Collaboration}, {Ade}, P.~A.~R., {Aghanim}, N., {et~al.}
  2016{\natexlab{b}},
  \href{http://dx.doi.org/10.1051/0004-6361/201424945}{\JournalTitle{\aap},
  586, A132}

\bibitem[{{Planck Collaboration} {et~al.}(2017){Planck Collaboration},
  {Aghanim}, {Ashdown}, {Aumont}, {Baccigalupi}, {Ballardini}, {Banday},
  {Barreiro}, {Bartolo}, {Basak}, {Benabed}, {Bernard}, {Bersanelli},
  {Bielewicz}, {Bonaldi}, {Bonavera}, {Bond}, {Borrill}, {Bouchet},
  {Boulanger}, {Bracco}, {Burigana}, {Calabrese}, {Cardoso}, {Chiang},
  {Colombo}, {Combet}, {Comis}, {Crill}, {Curto}, {Cuttaia}, {Davis}, {de
  Bernardis}, {de Rosa}, {de Zotti}, {Delabrouille}, {Delouis}, {Di Valentino},
  {Dickinson}, {Diego}, {Dor{\'e}}, {Douspis}, {Ducout}, {Dupac}, {Dusini},
  {Efstathiou}, {Elsner}, {En{\ss}lin}, {Eriksen}, {Falgarone}, {Fantaye},
  {Finelli}, {Frailis}, {Fraisse}, {Franceschi}, {Frolov}, {Galeotta}, {Galli},
  {Ganga}, {G{\'e}nova-Santos}, {Gerbino}, {Ghosh}, {Giard},
  {Gonz{\'a}lez-Nuevo}, {G{\'o}rski}, {Gregorio}, {Gruppuso}, {Gudmundsson},
  {Hansen}, {Helou}, {Herranz}, {Hivon}, {Huang}, {Jaffe}, {Jones},
  {Keih{\"a}nen}, {Keskitalo}, {Kisner}, {Krachmalnicoff}, {Kunz},
  {Kurki-Suonio}, {Lagache}, {L{\"a}hteenm{\"a}ki}, {Lamarre}, {Lasenby},
  {Lattanzi}, {Lawrence}, {Le Jeune}, {Levrier}, {Liguori}, {Lilje},
  {L{\'o}pez-Caniego}, {Lubin}, {Mac{\'{\i}}as-P{\'e}rez}, {Maggio}, {Maino},
  {Mandolesi}, {Mangilli}, {Maris}, {Martin}, {Mart{\'{\i}}nez-Gonz{\'a}lez},
  {Matarrese}, {Mauri}, {McEwen}, {Melchiorri}, {Mennella}, {Migliaccio},
  {Mitra}, {Miville-Desch{\^e}nes}, {Molinari}, {Moneti}, {Montier},
  {Morgante}, {Moss}, {Naselsky}, {N{\o}rgaard-Nielsen}, {Oxborrow}, {Pagano},
  {Paoletti}, {Partridge}, {Patrizii}, {Perdereau}, {Perotto}, {Pettorino},
  {Piacentini}, {Plaszczynski}, {Polenta}, {Puget}, {Rachen}, {Reinecke},
  {Remazeilles}, {Renzi}, {Rocha}, {Rossetti}, {Roudier},
  {Rubi{\~n}o-Mart{\'{\i}}n}, {Ruiz-Granados}, {Salvati}, {Sandri},
  {Savelainen}, {Scott}, {Sirignano}, {Sirri}, {Stanco}, {Suur-Uski}, {Tauber},
  {Tenti}, {Toffolatti}, {Tomasi}, {Tristram}, {Trombetti}, {Valiviita},
  {Vansyngel}, {Van Tent}, {Vielva}, {Wandelt}, {Wehus}, {Zacchei}, \&
  {Zonca}}]{PlanckL}
{Planck Collaboration}, {Aghanim}, N., {Ashdown}, M., {et~al.} 2017,
  \href{http://dx.doi.org/10.1051/0004-6361/201629164}{\JournalTitle{\aap},
  599, A51}

\bibitem[{{Rosolowsky} {et~al.}(2008){Rosolowsky}, {Pineda}, {Foster},
  {Borkin}, {Kauffmann}, {Caselli}, {Myers}, \& {Goodman}}]{Rosolowsky08a}
{Rosolowsky}, E.~W., {Pineda}, J.~E., {Foster}, J.~B., {et~al.} 2008,
  \href{http://dx.doi.org/10.1086/524299}{\JournalTitle{\apjs}, 175, 509}

\bibitem[{Schmidt {et~al.}(2008)Schmidt, Federrath, \&
  Klessen}]{PhysRevLett.101.194505}
Schmidt, W., Federrath, C., \& Klessen, R. 2008,
  \href{http://dx.doi.org/10.1103/PhysRevLett.101.194505}{\JournalTitle{Phys.
  Rev. Lett.}, 101, 194505}

\bibitem[{{Schmidt, W.} {et~al.}(2009){Schmidt, W.}, {Federrath, C.}, {Hupp,
  M.}, {Kern, S.}, \& {Niemeyer, J. C.}}]{10.1051/0004-6361}
{Schmidt, W.}, {Federrath, C.}, {Hupp, M.}, {Kern, S.}, \& {Niemeyer, J. C.}
  2009,
  \href{http://dx.doi.org/10.1051/0004-6361:200809967}{\JournalTitle{A\&A},
  494, 127}

\bibitem[{{Schneider} {et~al.}(2003){Schneider}, {Ferrara}, {Salvaterra},
  {Omukai}, \& {Bromm}}]{Schneider2003}
{Schneider}, R., {Ferrara}, A., {Salvaterra}, R., {Omukai}, K., \& {Bromm}, V.
  2003, \href{http://dx.doi.org/10.1038/nature01579}{\JournalTitle{\nat}, 422,
  869}

\bibitem[{{Slepian} \& {Eisenstein}(2015)}]{Slepian15_alg}
{Slepian}, Z., \& {Eisenstein}, D.~J. 2015,
  \href{http://dx.doi.org/10.1093/mnras/stv2119}{\JournalTitle{\mnras}, 454,
  4142}

\bibitem[{{Slepian} \& {Eisenstein}(2016)}]{Slepian16_alg_WFTs}
---. 2016,
  \href{http://dx.doi.org/10.1093/mnrasl/slv133}{\JournalTitle{\mnras}, 455,
  L31}

\bibitem[{{Slepian} \& {Eisenstein}(2017)}]{Slepian_aniso_alg17}
---. 2017, \JournalTitle{ArXiv e-prints},
  \href{http://arxiv.org/abs/1709.10150}{{\sffamily arXiv:1709.10150}}

\bibitem[{{Slepian} {et~al.}(2015){Slepian}, {Eisenstein}, {Beutler}, {Cuesta},
  {Ge}, {Gil-Mar{\'{\i}}n}, {Ho}, {Kitaura}, {McBride}, {Nichol}, {Percival},
  {Rodr{\'{\i}}guez-Torres}, {Ross}, {Scoccimarro}, {Seo}, {Tinker}, {Tojeiro},
  \& {Vargas-Maga{\~n}a}}]{Slepian16_comp_3PCF}
{Slepian}, Z., {Eisenstein}, D.~J., {Beutler}, F., {et~al.} 2015,
  \JournalTitle{ArXiv e-prints},
  \href{http://arxiv.org/abs/1512.02231}{{\sffamily arXiv:1512.02231}}

\bibitem[{{Slepian} {et~al.}(2016){Slepian}, {Eisenstein}, {Blazek},
  {Brownstein}, {Chuang}, {Gil-Mar{\'{\i}}n}, {Ho}, {Kitaura}, {McEwen},
  {Percival}, {Ross}, {Rossi}, {Seo}, {Slosar}, \&
  {Vargas-Maga{\~n}a}}]{Slepian16_RV_const}
{Slepian}, Z., {Eisenstein}, D.~J., {Blazek}, J.~A., {et~al.} 2016,
  \JournalTitle{ArXiv e-prints},
  \href{http://arxiv.org/abs/1607.06098}{{\sffamily arXiv:1607.06098}}

\bibitem[{{Slepian} {et~al.}(2017){Slepian}, {Eisenstein}, {Brownstein},
  {Chuang}, {Gil-Mar{\'{\i}}n}, {Ho}, {Kitaura}, {Percival}, {Ross}, {Rossi},
  {Seo}, {Slosar}, \& {Vargas-Maga{\~n}a}}]{Slepian17_BAO_3PCF}
{Slepian}, Z., {Eisenstein}, D.~J., {Brownstein}, J.~R., {et~al.} 2017,
  \href{http://dx.doi.org/10.1093/mnras/stx488}{\JournalTitle{\mnras}, 469,
  1738}

\bibitem[{{Stanimirovic} {et~al.}(1999){Stanimirovic}, {Staveley-Smith},
  {Dickey}, {Sault}, \& {Snowden}}]{Stanimirovic99a}
{Stanimirovic}, S., {Staveley-Smith}, L., {Dickey}, J.~M., {Sault}, R.~J., \&
  {Snowden}, S.~L. 1999, \JournalTitle{\mnras}, 302, 417

\bibitem[{{Stanimirovi{\'c}} {et~al.}(2004){Stanimirovi{\'c}},
  {Staveley-Smith}, \& {Jones}}]{Stanimirovic04a}
{Stanimirovi{\'c}}, S., {Staveley-Smith}, L., \& {Jones}, P.~A. 2004,
  \href{http://dx.doi.org/10.1086/381869}{\JournalTitle{\apj}, 604, 176}

\bibitem[{{Tofflemire} {et~al.}(2011){Tofflemire}, {Burkhart}, \&
  {Lazarian}}]{tofflemire11}
{Tofflemire}, B.~M., {Burkhart}, B., \& {Lazarian}, A. 2011,
  \href{http://dx.doi.org/10.1088/0004-637X/736/1/60}{\JournalTitle{\apj}, 736,
  60}

\bibitem[{Vazquez-Semadeni(1994)}]{Vazquez-Semadeni1994}
Vazquez-Semadeni, E. 1994,
  \href{http://dx.doi.org/10.1086/173847}{\JournalTitle{ApJ}, 423, 681}

\bibitem[{{Vazquez-Semadeni} {et~al.}(1997){Vazquez-Semadeni},
  {Ballesteros-Paredes}, \& {Rodriguez}}]{vazquezsemadeni97}
{Vazquez-Semadeni}, E., {Ballesteros-Paredes}, J., \& {Rodriguez}, L.~F. 1997,
  \href{http://dx.doi.org/10.1086/303432}{\JournalTitle{\apj}, 474, 292}

\end{thebibliography}

\end{document}